\documentclass{article}

\usepackage{arxiv}
\usepackage{subfigure}
\usepackage{comment}

\usepackage[utf8]{inputenc} 
\usepackage{CJKutf8}
\usepackage[T1]{fontenc}    
\usepackage{hyperref}       
\usepackage{url}            
\usepackage{booktabs}       
\usepackage{amsfonts}       
\usepackage{nicefrac}       
\usepackage{microtype}      
\usepackage{lipsum}
\usepackage{graphicx}
\graphicspath{ {./images/} }
\usepackage{enumitem}
\setlist{leftmargin=3.5mm}
\usepackage[toc, title, titletoc]{appendix}

\usepackage[normalem]{ulem}

\usepackage{xcolor}
\usepackage{soul}
\newcommand{\note}[1]{\textbf{\color{red}[***** #1 *****]}}

\title{Is Working From Home The New Norm? An Observational Study Based On A Large Geo-tagged COVID-19 Twitter Dataset}
\date{}

\author{
 Yunhe Feng \\
  Electrical Engineering \& Computer Science\\
  University of Tennessee\\
  Knoxville, TN 37996 \\
  \texttt{yunhefeng@utk.edu} \\
  \And
  Wenjun Zhou \\
  Business Analytics \& Statistics\\
  University of Tennessee\\
  Knoxville, TN 37996 \\
  \texttt{wzhou4@utk.edu} \\
}

\begin{document}
\maketitle

\begin{abstract}%
As the COVID-19 pandemic swept over the world, people discussed facts, expressed opinions, and shared sentiments on social media.
Since the reaction to COVID-19 in different locations may be tied to local cases, government regulations, healthcare resources and socioeconomic factors, 
we curated a large geo-tagged Twitter dataset and performed exploratory analysis by location.
Specifically, we collected 650,563 unique geo-tagged tweets across the United States (50 states and Washington, D.C.) covering the date range from January 25 to May 10, 2020.
Tweet locations enabled us to conduct region-specific studies such as tweeting volumes and sentiment, sometimes in response to local regulations and reported COVID-19 cases.
During this period, many people started working from home.
The gap between workdays and weekends in hourly tweet volumes inspired us to propose algorithms to estimate work engagement during the COVID-19 crisis.
This paper also summarizes themes and topics of tweets in our dataset using both social media exclusive tools (i.e., \#hashtags, @mentions) and the latent Dirichlet allocation model.
We welcome requests for data sharing and conversations for more insights.

\textbf{Dataset link:} \url{http://covid19research.site/geo-tagged_twitter_datasets/}

\end{abstract}


\section{Introduction}

The COVID-19 pandemic has had a widespread impact on people's daily lives all over the globe.
According to local pandemic conditions, countries worldwide adopted various containment policies to protect their residents and slow down the spread of COVID-19.
Although countries like Sweden and South Korea did not lock down cities during the pandemic, most of the other countries, including China, Italy, Spain, and India, imposed long and stringent lockdowns to restrict gathering and social contact.
Inside the same country, different strategies and timelines were also set by regions and cities to ``flatten the curve'' and fight against the COVID-19 crisis.
People 
expressed various opinions, attitudes, and emotions on the same COVID-19 regulations due to local hospital resources, economic statuses, demographics, and many other geographic factors.
Therefore, it is reasonable and necessary to consider the location information when investigating the public reactions to COVID-19. 

However, it is challenging to conduct such large-scale studies using traditional surveys and questionnaires.
First, regulations and policies proposed and enforced in different regions are time-sensitive and changeable, making it hard to determine when surveys to be conducted and which survey questions to be included.
For example, California and Tennessee implemented stay-at-home orders on different dates.
The initialized plannings and executive orders could also be tuned promptly, such as extending lockdowns due to the fast-growing COVID-19 confirmed cases.
Traditional surveys are not flexible enough for such changes.
Second, it is time-consuming and expensive to recruit a large number of participants to take surveys, because demographics (especially geographical locations) must be considered.
If a comparative spatial study is conducted, it takes more time to recruit participants from multiple regions.

In this paper, we built a large geo-tagged Twitter dataset enabling fine-grained investigations of the public reactions to the COVID-19 pandemic.
More than 170 million English COVID-19 related tweets were harvested from Jan. 25 to May 10, 2020, among which 650,563 geo-tagged tweets posted within the United States were selected.
We took the U.S. as an example to explore the public reactions in different regions because states in the U.S. determined when, how, and what policies and regulations were imposed independently.
We first presented an overview of both daily and hourly tweet distributions.
Then, state-level and county-level geographic patterns of COVID-19 tweets were illustrated.
We also proposed algorithms to evaluate work engagement by comparing tweeting behaviors on workdays and weekends.
In addition, we extracted the involved popular topics using both social media exclusive tools (i.e., \#hashtags and @mentions) and general topic models.
Finally, we analyzed public emotions using polarized words and facial emojis.

We summarized the contributions and findings of this paper as follows:
\begin{itemize}
    \item A large geo-tagged COVID-19 Twitter dataset, containing more than 650,000 tweets collected from Jan. 25 to May 10 2020 in the United States, was built and published at \url{http://covid19research.site/geo-tagged_twitter_datasets/}.
    We listed tweet IDs for all 50 states and Washington D.C. respectively.
    \item We profiled geospatial distributions of COVID-19 tweets at multiple location levels, and reported the difference between states after normalizing tweet volumes based on COVID-19 case and death numbers. For example, we found residents in Oregon, Montana, Texas, and California reacted more intensely to the confirmed cases and deaths than other states.
    \item We defined work engagement measurements based on the difference between workdays and weekends by hourly tweeting volumes. 
    \item 
    When studying work engagement patterns after lockdown and reopen, we reporeted a few interesting findings. For example, the New York state showed lower work engagement than other states in the first week under stay-at-home orders. The average hourly work engagement in the afternoon (i.e., from 13:00 to 16:59) in the first week of reopening was much higher than the first week of staying at home.
    \item We also conducted a comprehensive social sentiment analysis via facial emojis to measure the general public's emotions on stay-at-home orders, reopening, the first/hundredth/thousandth confirmed cases, and the first/hundredth/thousandth deaths. We observed that negative moods dominated the public sentiment over these key COVID-19 events, which showed a similar pattern across states.
\end{itemize}

\section{Temporal Patterns}

In this section, we first provide an overview of tweet daily distributions, demonstrating when COVID-19 tweets became viral.
Next, the hourly distributions during different periods were illustrated.
We then proposed methods to measure work engagement by comparing the hourly tweeting frequencies on workdays and weekends.
We also studied the influence of COVID-19 regulations, such as stay-at-home orders and reopening, on work engagement.

\subsection{Daily Patterns}

\begin{figure}[h]
  \includegraphics[width=\linewidth]{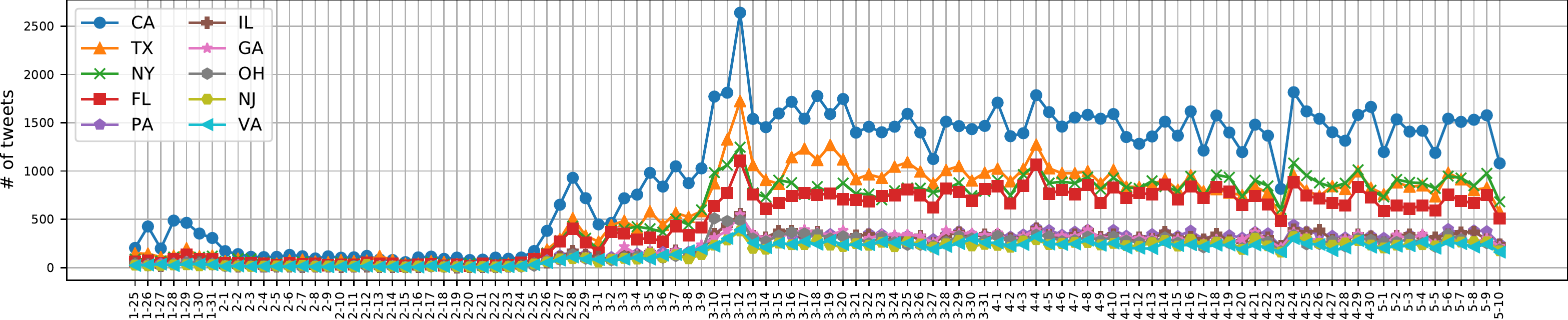}
  \caption{The daily number of tweets from the top 10 states generating most tweets. 
  \label{fig:daily_pattern_after_fixed}}
\end{figure}

Figure~\ref{fig:daily_pattern_after_fixed} shows the daily distribution of geo-tagged tweets within the top 10 states with the highest tweet volumes.\footnote{As mentioned in Appendix~\ref{sec:data_collection}, we lost around one-third detailed tweets between Mar. 18 and Apr. 4 due to the corrupted data.
But we recorded the daily tweet counts during this period (see the dashed lines in Figure~\ref{fig:daily_pattern}).
Our crawlers shut down for 8 hours and 9 hours On Mar. 27 and Apr. 23 respectively, which caused the data gaps in the two days.}
We can see that daily tweet volumes generated by different states show similar trends.
In fact, we tested statistical relationships regarding the daily volumes over time for all pairs of two arbitrary states, and found strong linear
correlations existed among 93.2\% state pairs with a Pearson's $r>0.8$ and $p < 0.001$.

Based on key dates, we split the entire observation period into the following three phases.
\begin{itemize}
    \item Phase 1 (from Jan. 25 to Feb. 24, 31 days): people mentioned little about COVID-19 except for a small peak at the end of January.
    \item Phase 2 (from Feb. 25 to Mar. 14, 19 days): the number of COVID-19 related tweets began to increase quickly. On Feb. 25 U.S. health officials warned the COVID-19 community spread in America was coming~\cite{feb25warning}.
    On March 13, the U.S. declared the national emergency due to COVID-19~\cite{mar13emergency}. 
    \item Phase 3 (from Mar. 15 to May 10, 57 days): people began to adjust to the new normal caused by COVID-19, such as working from home and city lockdowns.
\end{itemize}

\subsection{Hourly Patterns}
\begin{figure}[ht]
\subfigure[All
  \label{fig:distribution_by_hour}]{\includegraphics[width=0.24\linewidth]{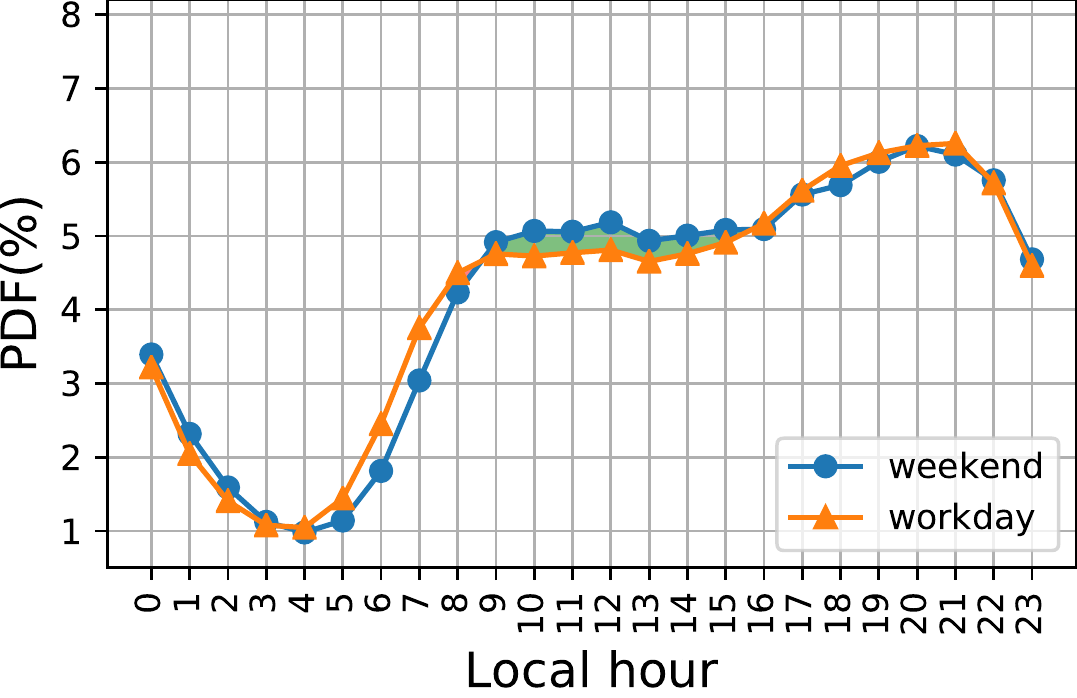}}
\subfigure[Phase 1
  \label{fig:distribution_by_hour_phase_1}]{\includegraphics[width=0.24\linewidth]{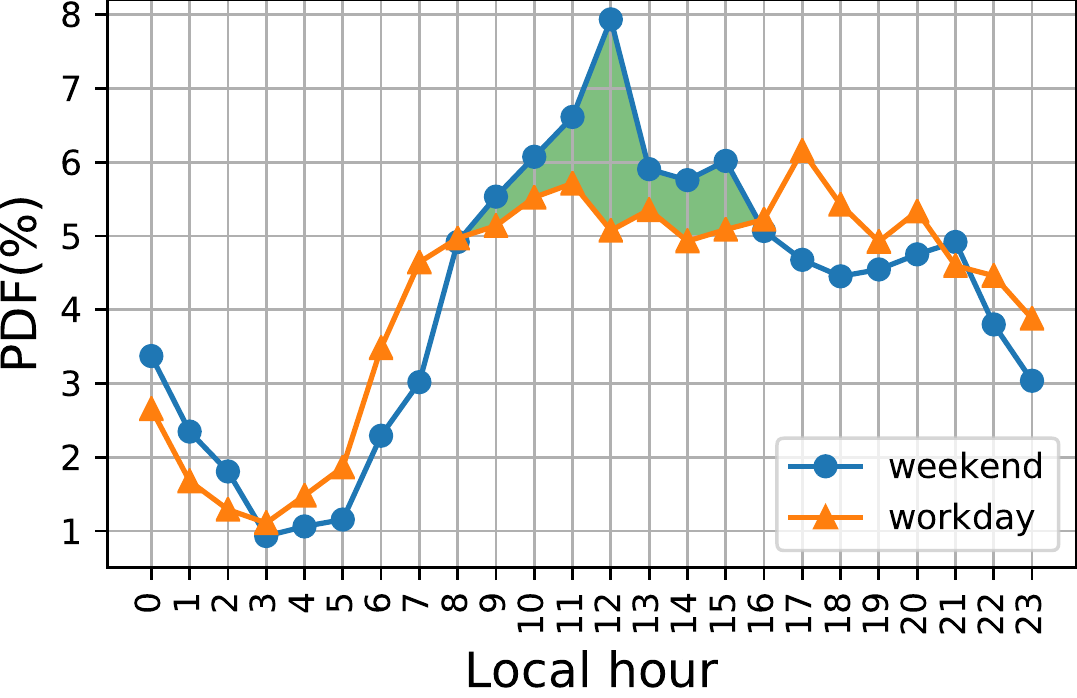}}
\subfigure[Phase 2
  \label{fig:distribution_by_hour_phase_2}]{\includegraphics[width=0.24\linewidth]{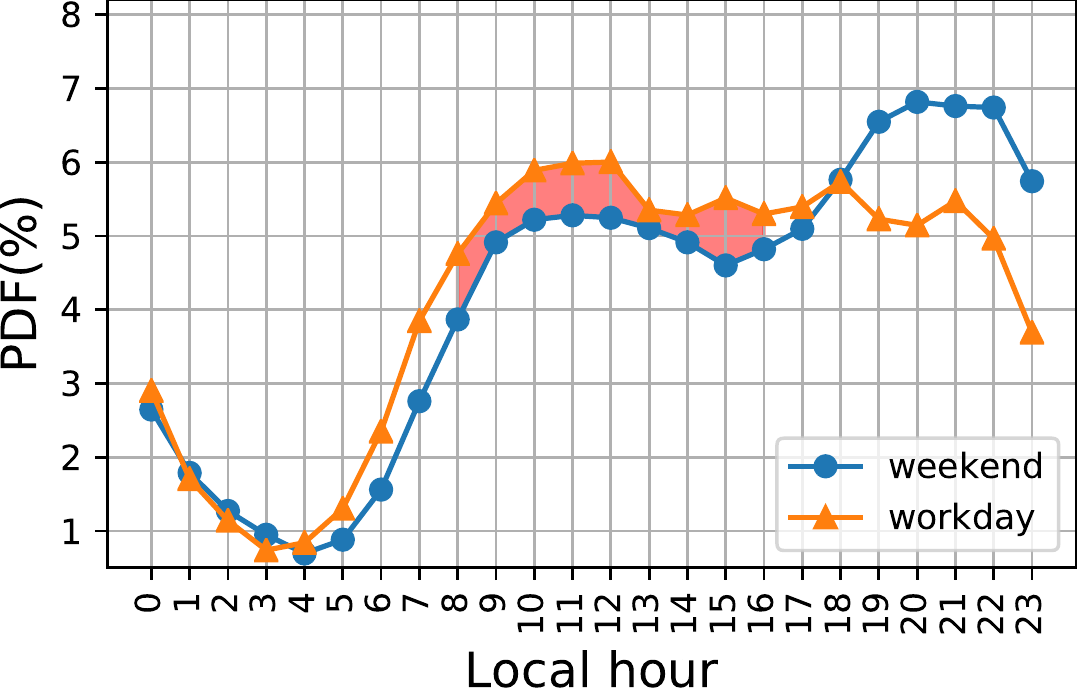}}
\subfigure[Phase 3
  \label{fig:distribution_by_hour_phase_3}]{\includegraphics[width=0.24\linewidth]{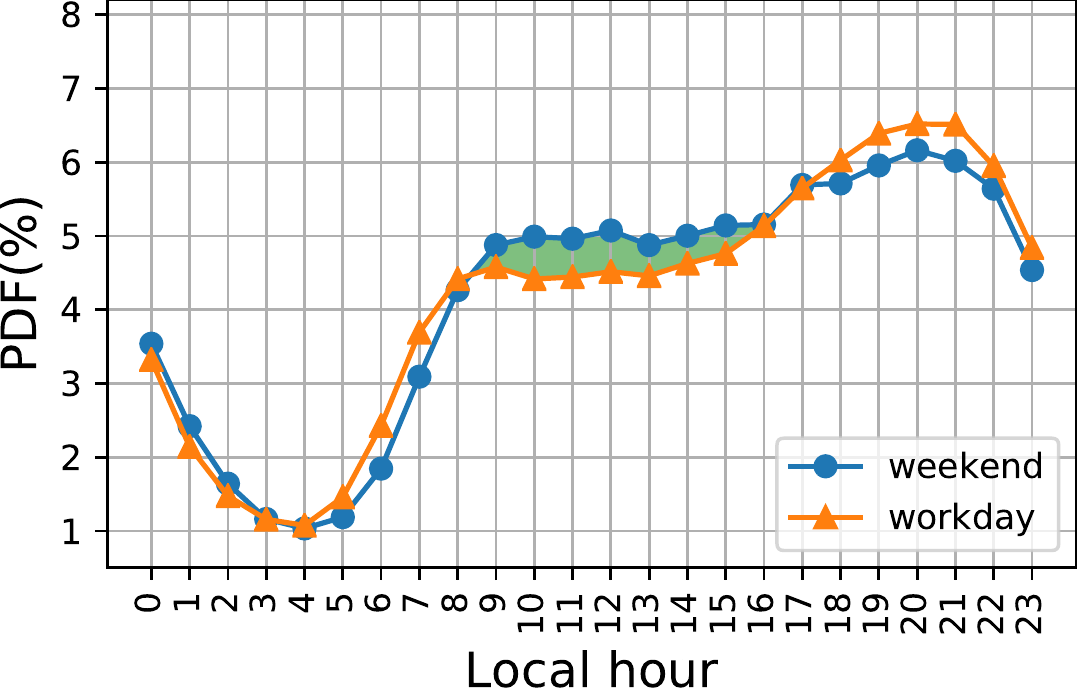}}
  \caption{Hourly distribution in three phases. The tweeting frequency gap during business hours are colored as green if the hourly frequency on weekends are higher than workdays. Otherwise, the gap is colored as red.
  \label{fig:hourly_distribution}}
\end{figure}

For each tweet, we converted the UTC time zone to its local time zone\footnote{For states spanning multiple time zones, we took the time zone covering most areas inside the state. For example, we used Eastern Standard Time (EST) when processing tweets from Michigan because EST is adopted by most of the state. Except for Arizona and Hawaii, we switched to Daylight Saving Time (DST) for all states after Mar. 8, 2020} according to the state where it was posted.
The aggregated hourly tweeting distributions in different phases are shown in Figure~\ref{fig:hourly_distribution}.
The tweeting behaviors on workdays and weekends were studied separately because we wanted to figure out how the working status impacted on tweeting patterns.
We colored the tweeting frequency gaps during business hours (8:00-16:59) as green if people tweeted more frequently on weekends than workdays.
Otherwise, the hourly gap is colored as red.

In Phase 1, there existed a tweeting gap from 8:00 to 16:59 between workdays and weekends.
The tweeting peak occurred at 12:00-12:59 on weekends but at 17:00-17:59 on workdays. 
We think it may be explained by the fact that people engage at work during regular working hours and have little time to post tweets on workdays.
But they become free to express concerns on COVID-19 on Twitter after work.

The hourly distribution patterns changed in Phase 2 when confirmed COVID-19 cases increased quickly in the United States.
People posted COVID-19 tweets more frequently during business hours than at the same time slots on weekends, indicating COVID-19 had drawn great attention of workers when they were working.  

It is interesting to note that a green tweeting gap from 8:00 to 16:59 reappeared in Phase 3 when most people had worked from home.
These findings motivated us to take advantage of the tweeting frequencies on workdays and weekends to estimate work engagement in the COVID-19 crisis (see Section~\ref{sec:work_engagement}).

\section{Geographic Patterns}

Twitter users can tag tweets with general locations (e.g. city, neighborhood) or exact GPS locations.
In this section, we utilized the two types of tweet locations to explore geographic patterns of COVID-19 tweets at state and county levels.

\subsection{State-level Distribution}

\begin{figure}[ht]
\subfigure[Tweeting percentage in each state 
  \label{fig:distribution_in_space_US_map}]{\includegraphics[width=0.49\linewidth]{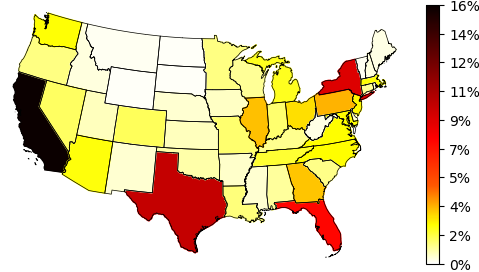}}
\subfigure[\# of geo-tagged tweets per 1000 residents in each state 
  \label{fig:distribution_in_space_US_map_normalized_by_popu}]{\includegraphics[width=0.49\linewidth]{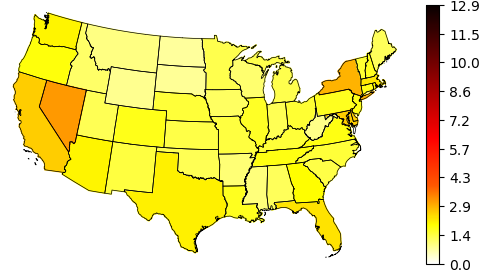}}
  
\subfigure[\# of geo-tagged tweets per COVID-19 case in each state 
  \label{fig:distribution_in_space_US_map_normalized_by_cases}]{\includegraphics[width=0.49\linewidth]{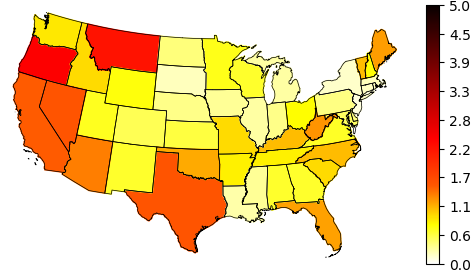}}
\subfigure[\# of geo-tagged tweets per COVID-19 death in each state 
  \label{fig:distribution_in_space_US_map_normalized_by_deaths}]{\includegraphics[width=0.49\linewidth]{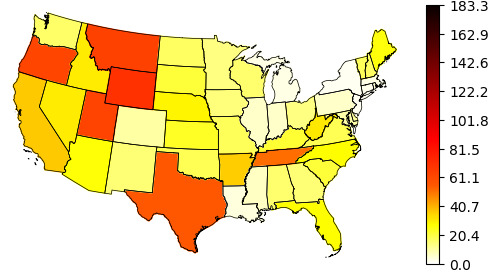}}  
  \caption{State-level geospatial distribution across the United States}
  \label{fig:distribution_in_space_US_map_all}
\end{figure}

We extracted the state information from both general and exact tweet locations and calculated tweet volume percentages for each state, as shown in Figure~\ref{fig:distribution_in_space_US_map}.
The most populated states, i.e., California, Texas, New York, and Florida, contributed the most tweets. In contrast, less populated states such as Wyoming, Montana, North Dakota, and South Dakota created the least tweets.
We measured the relationship between tweet volumes and populations for all states, and found a strong linear correlation existed (Pearson's $r=0.977$ and $p < 0.001$).

Then we normalized tweet volumes using state residential populations.
Figure~\ref{fig:distribution_in_space_US_map_normalized_by_popu} illustrates Washington D.C. posted the highest volume of tweets by every 1000 residents, followed by Nevada, New York, California, and Maryland. 
The rest states demonstrate similar patterns.
We think the top ranking of Washington D.C. might be caused by its functionality serving as a political center, where COVID-19 news and policies were spawned.

Unlike state populations, we did not find strong correlations between tweet counts and cumulative confirmed COVID-19 cases (Pearson's $r=0.544$ and $p < 0.001$) or deaths (Pearson's $r=0.450$ and $p < 0.001$).
We further normalized tweet volumes based on COVID-19 cumulative number of cases and deaths in each state.
Figure~\ref{fig:distribution_in_space_US_map_normalized_by_cases} and Figure~\ref{fig:distribution_in_space_US_map_normalized_by_deaths} shows the average number of tweets generated by each COVID-19 case and each death respectively.
Note that Hawaii and Alaska (not plotted in Figure~\ref{fig:distribution_in_space_US_map_normalized_by_cases} and Figure~\ref{fig:distribution_in_space_US_map_normalized_by_deaths}) ranked as the first and second in both scenarios.
Residents in states like Oregon, Montana, Texas, and California reacted sensitively to both confirmed cases and deaths, as these states dominated in Figure~\ref{fig:distribution_in_space_US_map_normalized_by_cases} and Figure~\ref{fig:distribution_in_space_US_map_normalized_by_deaths}.

\subsection{County-level Distribution}

\begin{figure}[ht]
\subfigure[Exact $(lat, lon)$ coordinates
  \label{fig:precise_geo_tagged_tweets}]{\includegraphics[width=0.49\linewidth]{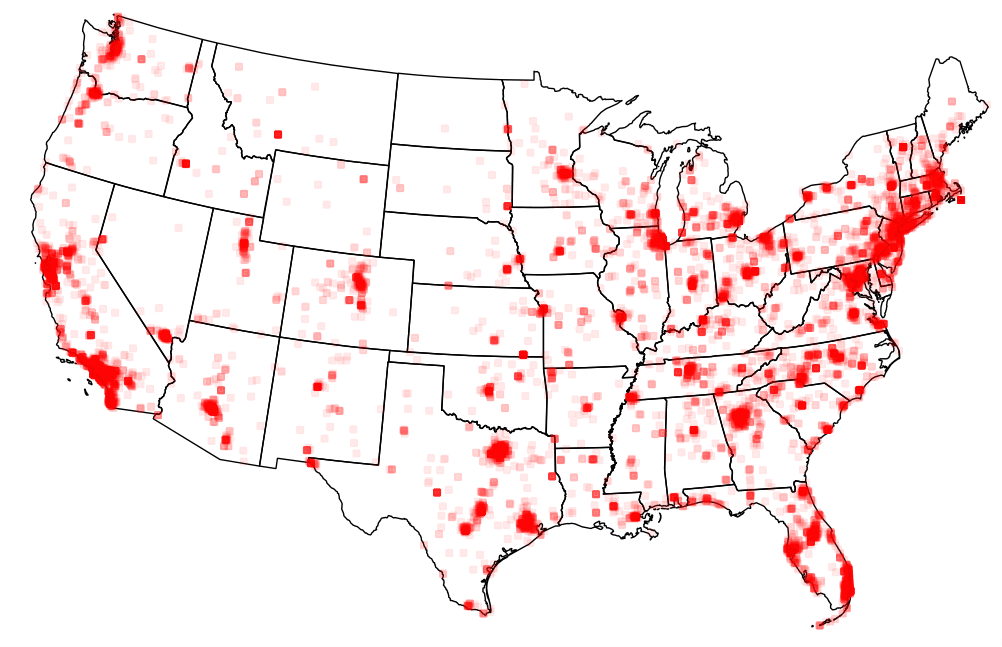}}
\subfigure[Geospatial distribution by county
  \label{fig:dict_parsed_US_coordinate}]{\includegraphics[width=0.49\linewidth]{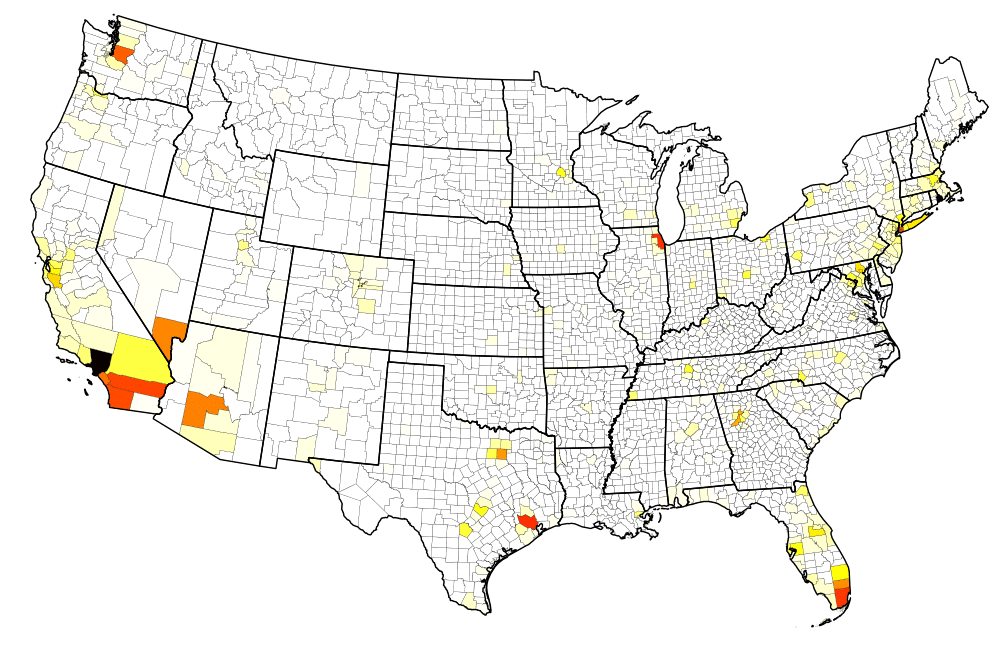}}
  \caption{Distribution of tweets tagged with exact GPS coordinates at the county level}
  \label{fig:city_distribution}
\end{figure}

We utilized GPS locations to profile the geographic distribution of COVID-19 tweets at the county level because general tweet locations might not contain county information. 
In our collected geo-tagged tweets, 3.95\% of them contained GPS locations.
We resorted to Nominatim~\cite{Nominatim}, a search engine for OpenStreetMap data, to identify the counties where each tweet GPS coordinate lay. 
Figure~\ref{fig:precise_geo_tagged_tweets} and Figure~\ref{fig:dict_parsed_US_coordinate} visualize the raw GPS coordinate and corresponding county distributions.
Large cities in each state demonstrated a higher tweeting density than small ones.
In fact, we found a strong correlations between GPS-tagged tweet counts and county populations (Pearson's $r=0.871$ and $p < 0.001$).
But such correlations did not hold true for cumulative confirmed COVID-19 cases (Pearson's $r=0.590$ and $p < 0.001$) or deaths (Pearson's $r=0.497$ and $p < 0.001$).


\section{Work Engagement Analysis}
\label{sec:work_engagement}

In this section, we first propose methods to measure hourly and daily work engagement. 
Then, we investigate how stay-at-home orders and reopening influenced hourly and daily work engagement respectively.
Note that we use the term of ``lockdown'' referring stay-at-home orders in this section.
The lockdown dates and reopening dates for each state are retrieved from Wikipedia~\cite{lockdownWiki} and New York Times~\cite{ReopenNYT} respectively. 

\subsection{The Work Engagement Index}

We assumed that (1) people would tweet less frequently during working hours if they engaged more on their working tasks;
(2) if people spent no time on work tasks during business hours on workdays, their tweeting behaviors kept the same as that on weekends, especially when people were confined at home.
We think the two assumptions are intuitive and reasonable as both Phase 1 and Phase 3 showed the meaningful working-hour tweeting gaps in Figure~\ref{fig:distribution_by_hour_phase_1} and Figure~\ref{fig:distribution_by_hour_phase_3}.

We took the tweeting gap size as an indicator of work engagement.
More specifically, let $h_i^j$ denote the tweet volume at the $i$-th hour on the $j$-th day in a week. 
For example, $h_8^2$ meant the number of tweets posted from 8:00 to 8:59 on Tuesdays (we took Monday as the first day in one week).
Accordingly, the total number of tweets on the $j$-th day in a week was represented by $T_j = \sum_{i=1}^{24}h_i^j$.
The total tweet volumes on workdays and weekends can be expressed as $T_{workday} = \sum_{j=1}^{5}T_j$ and $T_{weekend}  = \sum_{j=6}^{7}T_j$.

Note that we estimated both hourly and daily work engagement by considering at least seven days, because the data sparsity would lead to unreliable results if fewer days were involved.
The work engagement at $i$-th hour $H(i)$ could be defined as the ratio of the normalized tweeting frequency at $i$-th hour on weekends over that on workdays and minus one, as expressed in Equation~\ref{eqn:hourly_engagement}.

\begin{equation}\label{eqn:hourly_engagement}
    H(i) = \frac{\sum_{j=6}^{7}h_i^j}{T_{weekend} } / \frac{\sum_{j=1}^{5}h_i^j}{T_{workday} } - 1,
\end{equation}

where $i \in \{8, 10, 11, 12, 13, 14, 15, 16\}$.
A larger positive $H(i)$ indicates higher work engagement.
When $H(i)$ equals 0, it means there exists no difference on work engagement at $i$-th hour between workdays and weekends.
A positive value of $H(i)$ implies people are more engaged at work on workdays than weekends.
Although it is rare, a negative $H(i)$ means people fail to focus more on their work on workdays than weekends.


We also defined the daily (from Monday to Friday only) work engagement by aggregating tweeting frequencies in regular working hours (from 8:00 to 16:59).
The daily work engagement on $j$-th day was expressed as:
\begin{equation}
    D(j) = \frac{\sum_{j=6}^{7}\sum_{i=8}^{16}h_i^j}{T_{weekend} } / \frac{\sum_{i=8}^{16}h_i^j}{T_j} - 1
\end{equation}
where $j \in \{1, 2, 3, 4, 5\}$ and $T_j$ was the total tweet count on the $j$-th day.
Similar to hourly engagement, a larger positive $D(j)$ means higher work engagement.

\subsubsection{Stay-at-home Order Impacts on Work Engagement}
\label{sec:stay-at-home-work-engagement}

We chose the ten states that generated the most massive tweet volumes from 8:00 to 16:59 on each day of the first week after stay-at-home orders were enforced, to study the hourly and daily work engagement.
Table~\ref{tab:hourly_stay_at_home} illustrates the hourly work engagement of the ten states. 
Except for California and New York, all other eight states had positive average work engagement scores (see the second last column in Table~\ref{tab:hourly_stay_at_home}), implying people worked more extensively on workdays than weekends.
Georgia and Maryland demonstrated relatively higher average work engagements ($>0.15$).
Across all the ten states, people focused more on work tasks at 10:00 and 13:00 than other hour slots, and reached the lowest engagement score at 11:00 (see the second last row in Table~\ref{tab:hourly_stay_at_home}).

\begin{table}[h]
\setlength{\tabcolsep}{4pt}
\centering
\caption{Hourly work engagement scores (first stay-at-home week of each state)}
\begin{tabular}{llr|rrrrrrrrr|rr}
\toprule
\textbf{State} & \textbf{Date} & \textbf{\#Tweets} & \textbf{8:00} & \textbf{9:00} & \textbf{10:00} & \textbf{11:00} & \textbf{12:00} & \textbf{13:00} & \textbf{14:00} & \textbf{15:00} & \textbf{16:00} & \textbf{Avg.} & \textbf{Std.} \\ \midrule
\textbf{CA}    & Mar 19        & 8,091              & -0.016        & -0.063        & 0.206          & 0.081          & -0.018         & 0.136          & -0.178         & -0.184         & 0.009          & -0.003        & 0.131         \\
\textbf{TX}    & Apr 2         & 6,758              & -0.109        & 0.083         & 0.195          & 0.104          & 0.272          & 0.256          & -0.04          & 0.047          & 0.099          & 0.101         & 0.127         \\
\textbf{FL}    & Apr 3         & 5,582              & -0.123        & 0.057         & 0.367          & -0.092         & 0.209          & 0.115          & 0.279          & 0.097          & 0.374          & 0.143         & 0.181         \\
\textbf{NY}    & Mar 22        & 4,213              & -0.366        & 0.025         & 0.065          & 0.164          & 0.174          & 0.157          & -0.241         & -0.056         & -0.281         & -0.040         & 0.208         \\
\textbf{GA}    & Apr 3         & 2,334              & -0.226        & -0.108        & 0.328          & -0.347         & 0.053          & 0.388          & 0.831          & 0.473          & 0.000            & 0.155         & 0.378         \\
\textbf{PA}    & Apr 1         & 2,327              & -0.078        & 0.368         & 0.144          & 0.201          & -0.098         & 0.650           & -0.214         & 0.092          & 0.094          & 0.129         & 0.262         \\
\textbf{IL}    & Mar 21        & 1,639              & 0.337         & -0.096        & 0.321          & -0.296         & 0.252          & -0.038         & 0.242          & -0.089         & 0.033          & 0.074         & 0.223         \\
\textbf{MD}    & Mar 30        & 1,598              & 0.151         & 0.266         & 1.275          & -0.030          & -0.336         & -0.036         & -0.073         & 0.455          & 0.767          & 0.271         & 0.497         \\
\textbf{VA}    & Mar 30        & 1,595              & 0.460          & 0.065         & 0.792          & -0.200           & 0.024          & 0.135          & -0.240          & 0.182          & -0.089         & 0.125         & 0.328         \\
\textbf{AZ}    & Mar 31        & 1,508              & 0.381         & -0.097        & -0.160          & -0.020          & -0.038         & 0.624          & -0.185         & 0.674          & 0.054          & 0.137         & 0.334         \\ \midrule
\textbf{Avg.}  &               &                   & 0.041         & 0.050          & 0.353          & -0.043         & 0.049          & 0.239          & 0.018          & 0.169          & 0.106          &               &               \\
\textbf{Std.}  &               &                   & 0.278         & 0.160          & 0.406          & 0.190           & 0.187          & 0.244          & 0.343          & 0.278          & 0.284          &               &               \\ \bottomrule
\end{tabular}
\label{tab:hourly_stay_at_home}
\end{table}

Table~\ref{tab:daily_stay_at_home} shows the daily work engagement in the first week after stay-at-home orders were announced.
The average daily patterns for each state is very similar to the hourly ones.
For example, both the daily and hourly average engagements in California and New York were negative.
Based on average daily work engagement of the ten states, we found people put themselves more in their work on Thursday and Friday than Monday, Tuesday, and Wednesday (see the second last row in Table~\ref{tab:daily_stay_at_home}).  

\begin{table}[h]
\setlength{\tabcolsep}{5pt}
\centering
\caption{Daily work engagement scores (first stay-at-home week of each state)}
\begin{tabular}{llr|rrrrr|rr}
\toprule
\textbf{State} & \textbf{Date} & \textbf{\#Tweets} & \textbf{Mon.} & \textbf{Tue.} & \textbf{Wed.} & \textbf{Thu.} & \textbf{Fri.} & \textbf{Avg.} & \textbf{Std.} \\ \midrule
\textbf{CA}    & Mar 19        & 8,091              & -0.005        & -0.042        & -0.032        & 0.026         & -0.001        & -0.011        & 0.027         \\
\textbf{TX}    & Apr 2         & 6,758              & 0.097         & 0.045         & 0.001         & 0.383         & 0.083         & 0.122         & 0.151         \\
\textbf{FL}    & Apr 3         & 5,582              & 0.004         & 0.093         & 0.093         & 0.352         & 0.260          & 0.160          & 0.142         \\
\textbf{NY}    & Mar 22        & 4,213              & 0.036         & -0.103        & -0.07         & 0.076         & -0.115        & -0.035        & 0.086         \\
\textbf{GA}    & Apr 3         & 2,334              & 0.103         & 0.027         & 0.009         & 0.388         & 0.138         & 0.133         & 0.152         \\
\textbf{PA}    & Apr 1         & 2,327              & 0.115         & 0.013         & 0.009         & 0.336         & 0.201         & 0.135         & 0.138         \\
\textbf{IL}    & Mar 21        & 1,639              & 0.020          & 0.021         & 0.112         & 0.088         & -0.009        & 0.046         & 0.051         \\
\textbf{MD}    & Mar 30        & 1,598              & -0.047        & 0.311         & 0.117         & 0.410          & 0.668         & 0.292         & 0.274         \\
\textbf{VA}    & Mar 30        & 1,595              & -0.097        & 0.190          & 0.127         & 0.112         & 0.124         & 0.091         & 0.110          \\
\textbf{AZ}    & Mar 31        & 1,508              & 0.291         & 0.051         & 0.140          & 0.008         & 0.103         & 0.119         & 0.109         \\ \midrule
\textbf{Avg.}  &               &                   & 0.052         & 0.061         & 0.051         & 0.218         & 0.145         &               &               \\
\textbf{Std.}  &               &                   & 0.108         & 0.117         & 0.075         & 0.168         & 0.213         &               &               \\ \bottomrule
\end{tabular}
\label{tab:daily_stay_at_home}
\end{table}

\begin{figure}[h]
\subfigure[Average hourly work engagement per week
  \label{fig:avg_weekly_engagement_by_hour}]{\includegraphics[width=0.49\linewidth]{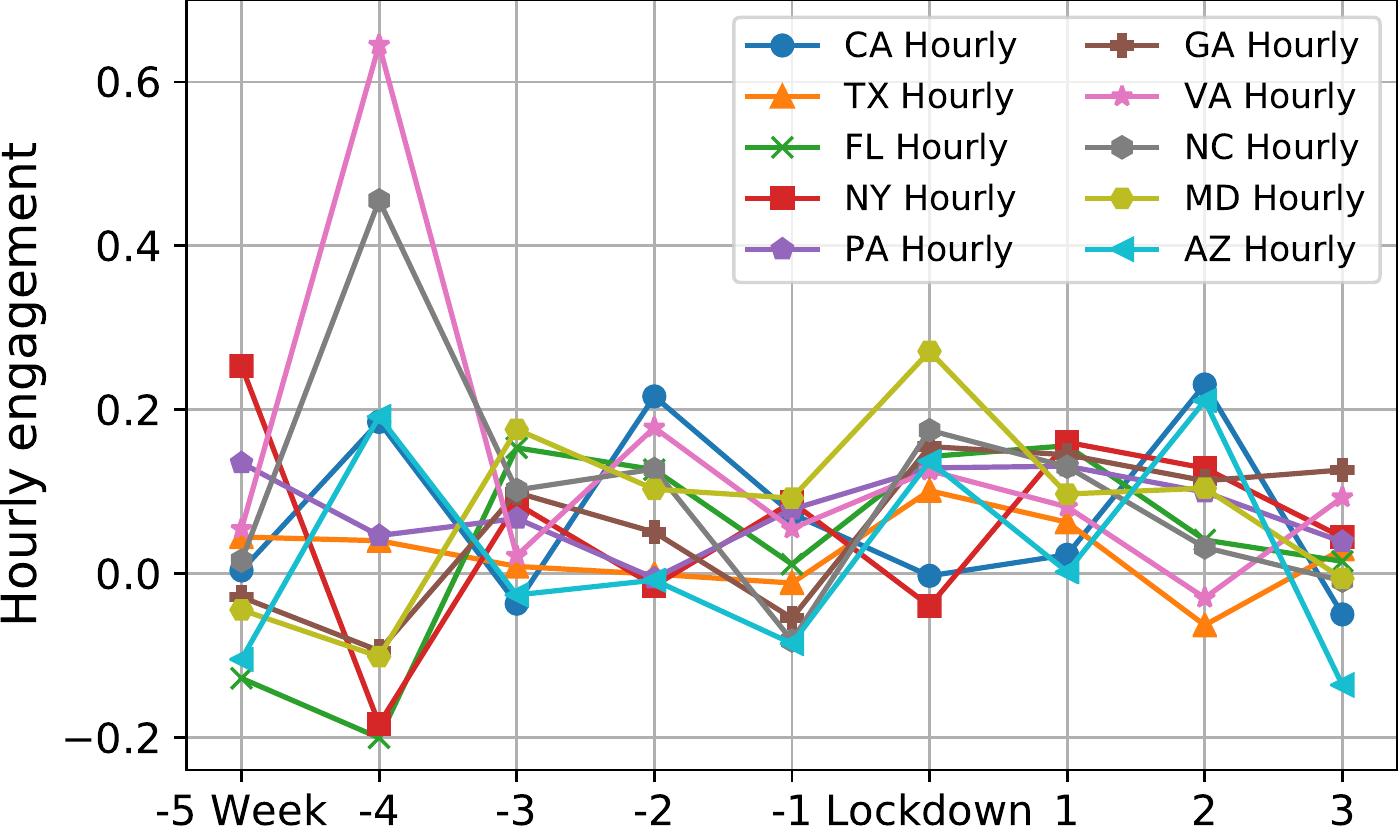}}
\subfigure[Average daily work engagement per week
  \label{fig:avg_weekly_engagement_by_day}]{\includegraphics[width=0.49\linewidth]{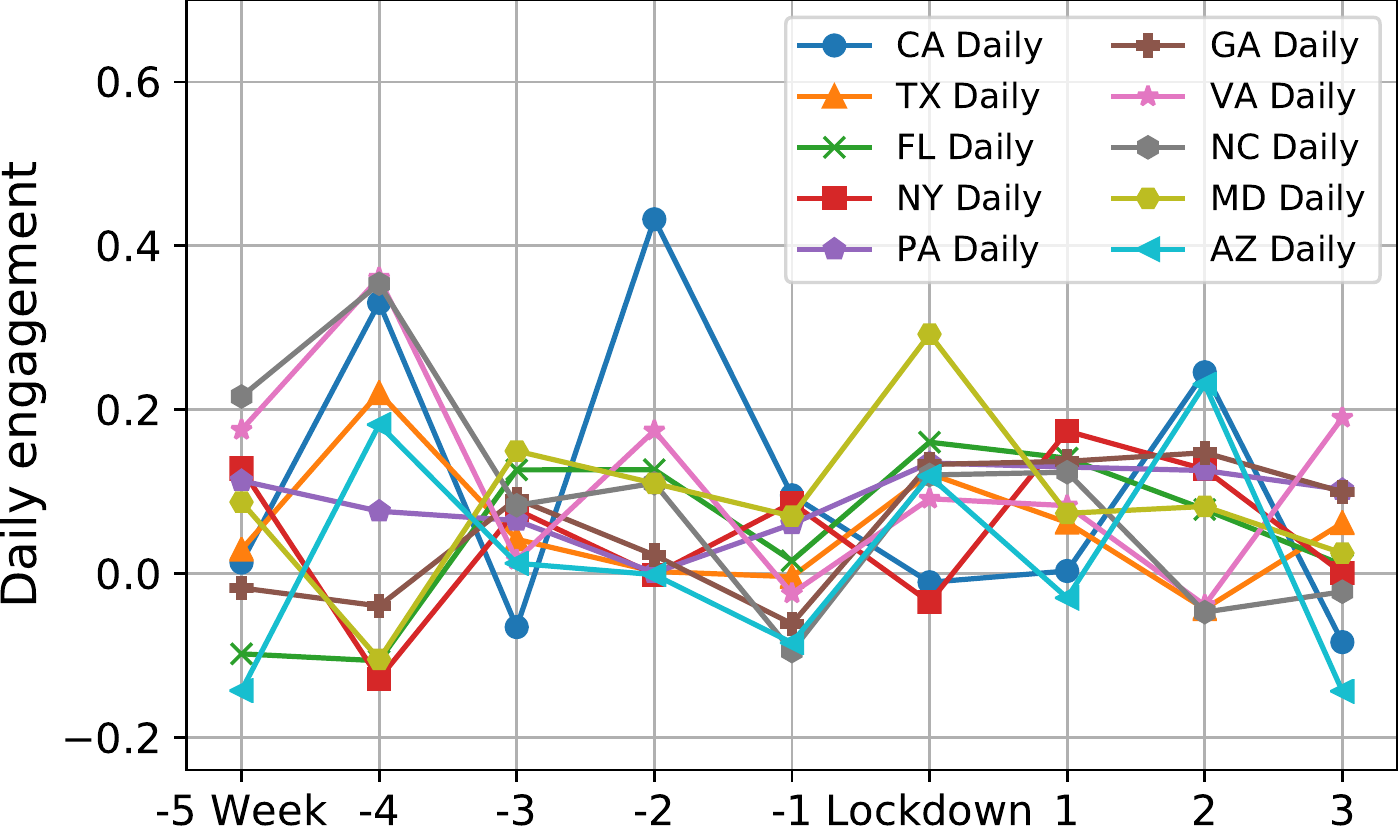}}
  \caption{Average hourly and daily work engagement in the first five weeks before and three weeks after local stay-at-home orders were released}
  \label{fig:weekly_engagement}
\end{figure}

Besides the first stay-at-home week, we reported the average hourly and daily work engagement for states in a more extended period ranging from five weeks ahead of and three weeks after stay-at-home orders were issued.
As shown in Figure~\ref{fig:weekly_engagement}, hourly and daily work engagement patterns of the same state are very similar along the nine weeks.
States performed very differently one month before local lockdowns (see x-axis=-4 and x-axis=-5 in Figure~\ref{fig:avg_weekly_engagement_by_hour} and Figure~\ref{fig:avg_weekly_engagement_by_day}).
Surprisingly, most states achieved higher work engagement in the first two weeks of lockdowns (see x-axis=Lockdown and x-axis=1 in Figure~\ref{fig:avg_weekly_engagement_by_hour} and Figure~\ref{fig:avg_weekly_engagement_by_day}) than before lockdowns.

\subsubsection{Reopening Impacts on Work Engagement}
\label{sec:stay-at-home-work-engagement}

\begin{figure}[ht]
\subfigure[Average hourly work engagement in $1^{st}$ week
  \label{fig:bar_hourly_engagment_lockdown_vs_reopening}]{\includegraphics[width=0.6\linewidth]{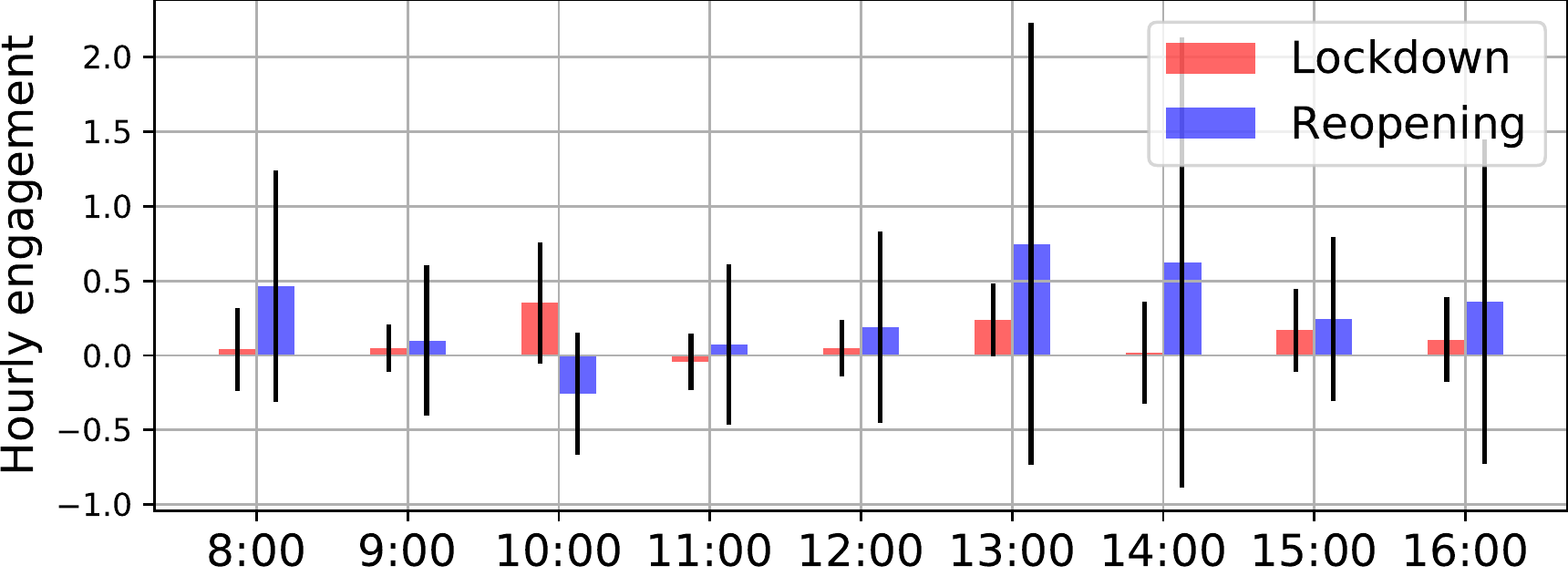}}
\subfigure[Average daily work engagement in $1^{st}$ week
  \label{fig:bar_daily_engagment_lockdown_vs_reopening}]{\includegraphics[width=0.37\linewidth]{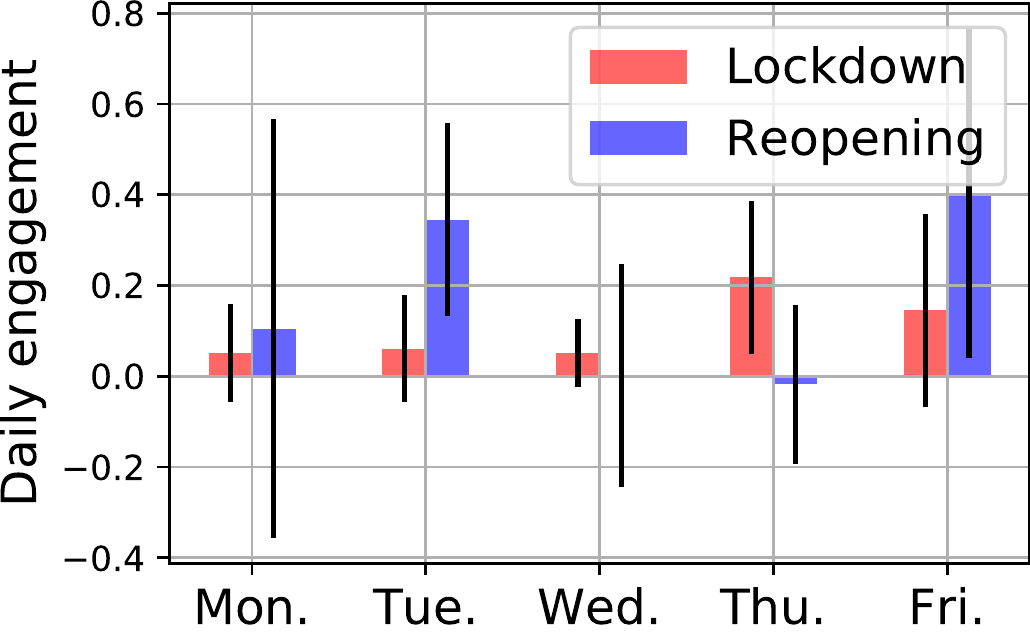}}
  \caption{Average hourly and daily work engagement in the first week of lockdowns and reopening}
  \label{fig:lockdown_vs_reopening}
\end{figure}

Some states had started to reopen partially since the end of April.
We selected the states that were partially reopened before May 3\footnote{We made sure each reopened state had at least seven-day tweets after its reopening in our dataset (Jan.25 - May 10).} to investigate their hourly and daily work engagement in the first week of reopening.
As Table~\ref{tab:hourly_reopening} and Table~\ref{tab:daily_reopening} show, averaged hourly and daily work engagement of the nine states except Alaska are positive.
People demonstrated much higher work engagement in the afternoon than in the morning (see the last second row in Table~\ref{tab:hourly_reopening}).
Figure~\ref{fig:lockdown_vs_reopening} demonstrates the afternoon work engagement of reopening is much larger than its counterpart in the first week of lockdowns.
Also, the average work engagement of reopening on Tuesday and Friday improves a lot when comparing with lockdowns.

\begin{table}[h]
\setlength{\tabcolsep}{4pt}
\centering
\caption{Hourly work engagement in the first week after reopen}
\begin{tabular}{llr|rrrrrrrrr|rr}
\toprule
\textbf{State} & \textbf{Date} & \textbf{\#Tweets} & \textbf{8:00} & \textbf{9:00} & \textbf{10:00} & \textbf{11:00} & \textbf{12:00} & \textbf{13:00} & \textbf{14:00} & \textbf{15:00} & \textbf{16:00} & \textbf{Avg.} & \textbf{Std.} \\ \midrule
\textbf{TX}    & May 1         & 5,927              & -0.308        & 0.215         & -0.187         & 0.010           & 0.170           & 0.125          & 0.913          & 0.367          & -0.098         & 0.134         & 0.360          \\
\textbf{GA}    & May 1         & 2,049              & -0.294        & 0.068         & 0.231          & 0.015          & -0.011         & 1.105          & 0.140           & 0.567          & 0.753          & 0.286         & 0.438         \\
\textbf{TN}    & May 1         & 1,280              & 0.236         & -0.119        & -0.116         & 0.242          & -0.129         & 0.105          & 1.895          & 0.535          & -0.061         & 0.288         & 0.643         \\
\textbf{CO}    & Apr 27        & 990               & -0.316        & 0.434         & 0.092          & 0.703          & -0.206         & 0.300            & -0.230          & -0.133         & 0.021          & 0.074         & 0.344         \\
\textbf{AL}    & May 1         & 603               & 1.453         & -0.328        & -0.398         & 0.472          & -0.097         & 2.753          & 3.047          & 0.104          & -0.146         & 0.762         & 1.336         \\
\textbf{MS}    & Apr 28        & 317               & 0.204         & -0.518        & -0.037         & -0.484         & -0.259         & 1.108          & 2.372          & 1.409          & 3.014          & 0.757         & 1.293         \\
\textbf{ID}    & May 1         & 184               & 0.523         & 0.692         & -0.805         & 0.587          & 0.523          & 3.231          & -1.000           & -0.154         & -0.683         & 0.324         & 1.275         \\
\textbf{AK}    & Apr 25        & 142               & 0.898         & 0.898         & -1.000           & -1.000           & -0.051         & -1.000           & -0.526         & -0.431         & -0.209         & -0.269        & 0.748         \\
\textbf{MT}    & Apr 27        & 106               & 1.786         & -0.443        & -0.071         & 0.114          & 1.786          & -1.000           & -1.000           & -0.071         & 0.671          & 0.197         & 1.043         \\ \midrule
\textbf{Avg.}  &               &                   & 0.465         & 0.100           & -0.255         & 0.073          & 0.192          & 0.747          & 0.623          & 0.244          & 0.362          &               &               \\
\textbf{Std.}  &               &                   & 0.776         & 0.503         & 0.409          & 0.538          & 0.642          & 1.481          & 1.508          & 0.552          & 1.088          &               &               \\ \bottomrule
\end{tabular}
\label{tab:hourly_reopening}
\end{table}

\begin{table}[h]
\setlength{\tabcolsep}{5pt}
\centering
\caption{Daily work engagement in the first week of reopening}
\begin{tabular}{llr|rrrrr|rr}
\toprule
\textbf{State} & \textbf{Date} & \textbf{\#Tweets} & \textbf{Mon.} & \textbf{Tue.} & \textbf{Wed.} & \textbf{Thu.} & \textbf{Fri.} & \textbf{Avg.} & \textbf{Std.} \\ \midrule
\textbf{TX}    & May 1         & 5,927              & 0.053         & 0.196         & -0.053        & 0.013         & 0.521         & 0.146         & 0.229         \\
\textbf{GA}    & May 1         & 2,049              & 0.099         & 0.362         & 0.168         & 0.158         & 0.504         & 0.258         & 0.169         \\
\textbf{TN}    & May 1         & 1,280              & 0.016         & 0.324         & 0.020          & -0.019        & 0.958         & 0.260          & 0.414         \\
\textbf{CO}    & Apr 27        & 990               & -0.213        & 0.334         & -0.077        & 0.049         & 0.502         & 0.119         & 0.294         \\
\textbf{AL}    & May 1         & 603               & 0.321         & 0.367         & 0.230          & 0.062         & 0.660          & 0.328         & 0.219         \\
\textbf{MS}    & Apr 28        & 317               & 0.310          & 0.787         & 0.239         & 0.226         & 0.245         & 0.361         & 0.240          \\
\textbf{ID}    & May 1         & 184               & 1.110          & 0.327         & -0.282        & -0.231        & 0.108         & 0.206         & 0.564         \\
\textbf{AK}    & Apr 25        & 142               & -0.486        & -0.020         & -0.449        & -0.327        & -0.327        & -0.322        & 0.183         \\
\textbf{MT}    & Apr 27        & 106               & -0.265        & 0.429         & 0.224         & -0.095        & 0.457         & 0.150          & 0.320          \\  \midrule
\textbf{Avg.}  &               &                   & 0.105         & 0.345         & 0.002         & -0.018        & 0.403         &               &               \\
\textbf{Std.}  &               &                   & 0.461         & 0.212         & 0.245         & 0.176         & 0.363         &               &               \\ \bottomrule
\end{tabular}
\label{tab:daily_reopening}
\end{table}

\section{Content Analysis}
In this section, we summarized and revealed the themes people discussed on Twitter.
Social network exclusive tools (i.e., \#hashtags and @mentions) and general text-based topic models were used to infer underlying tweet topics during the COVID-19 pandemic.

\subsection{Top Hashtags}
\begin{figure}[h]
\centering
  \includegraphics[width=\linewidth]{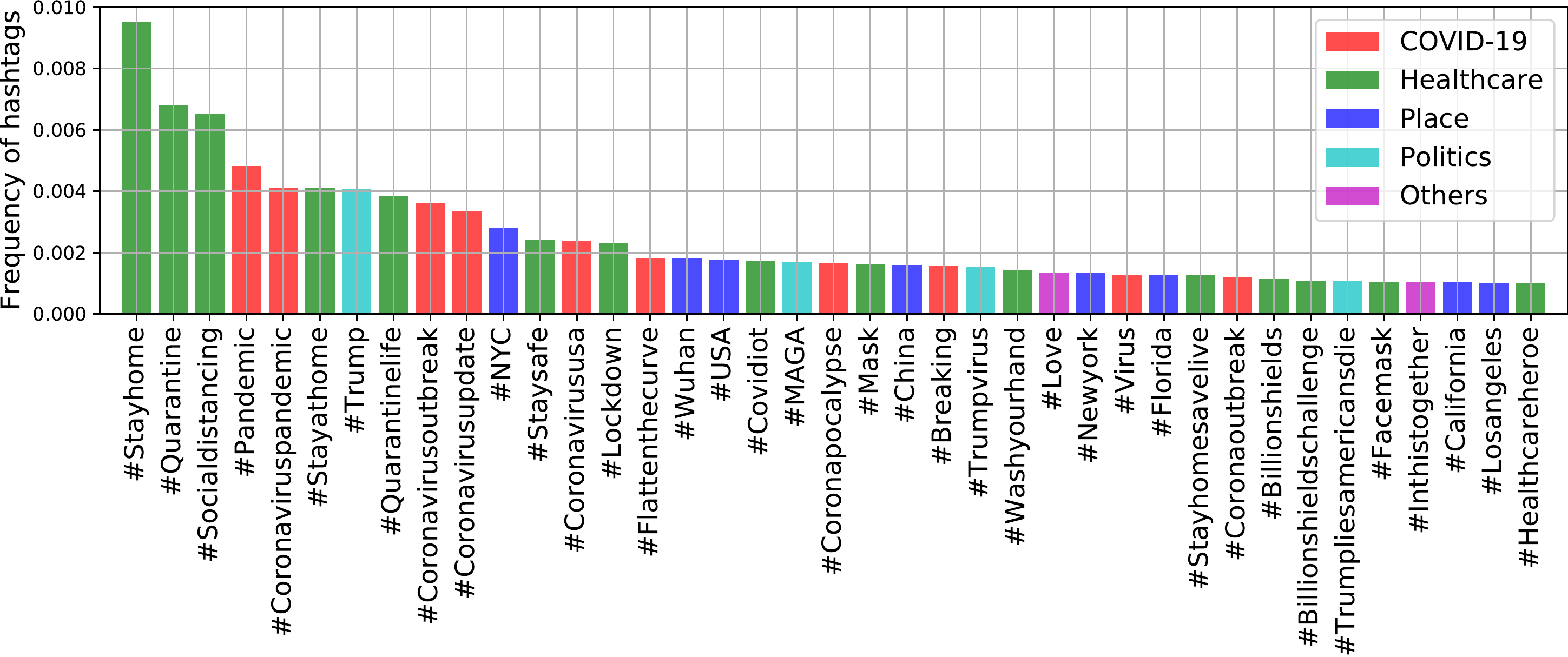}
  \caption{Top 40 most popular \#hashtags. \#COVID-19 and its variations (accounting for more than 25\%) are not plotted. 
  \label{fig:hashtags}}
\end{figure}

Hashtags are widely used on social networks to categorize topics and increase engagement.
According to Twitter, hashtagged words that become very popular are often trending topics.
We found \#hashtags were extensively used in geo-tagged COVID-19 tweets --- each tweet contained 0.68 \#hashtags on average.
Our dataset covered more than 86,000 unique \#hashtags, and 95.2\% of them appeared less than 10 times.
\#COVID-19 and its variations (e.g., ``Covid\_19'', and ``Coronavid19'') were the most popular ones, accounting for over 25\% of all \#hashtags.
To make the visualization of \#hashtags more readable, we did not plot \#COVID-19 and its variations in Figure~\ref{fig:hashtags}.
In other words, Figure~\ref{fig:hashtags} displays top \#hashtags starting from the second most popular one.
All \#hashtags were grouped into five categories, namely COVID-19, Healthcare, Place, Politics, and Others.

\subsection{Top Mentions}
\begin{figure}[h]
\centering
  \includegraphics[width=\linewidth]{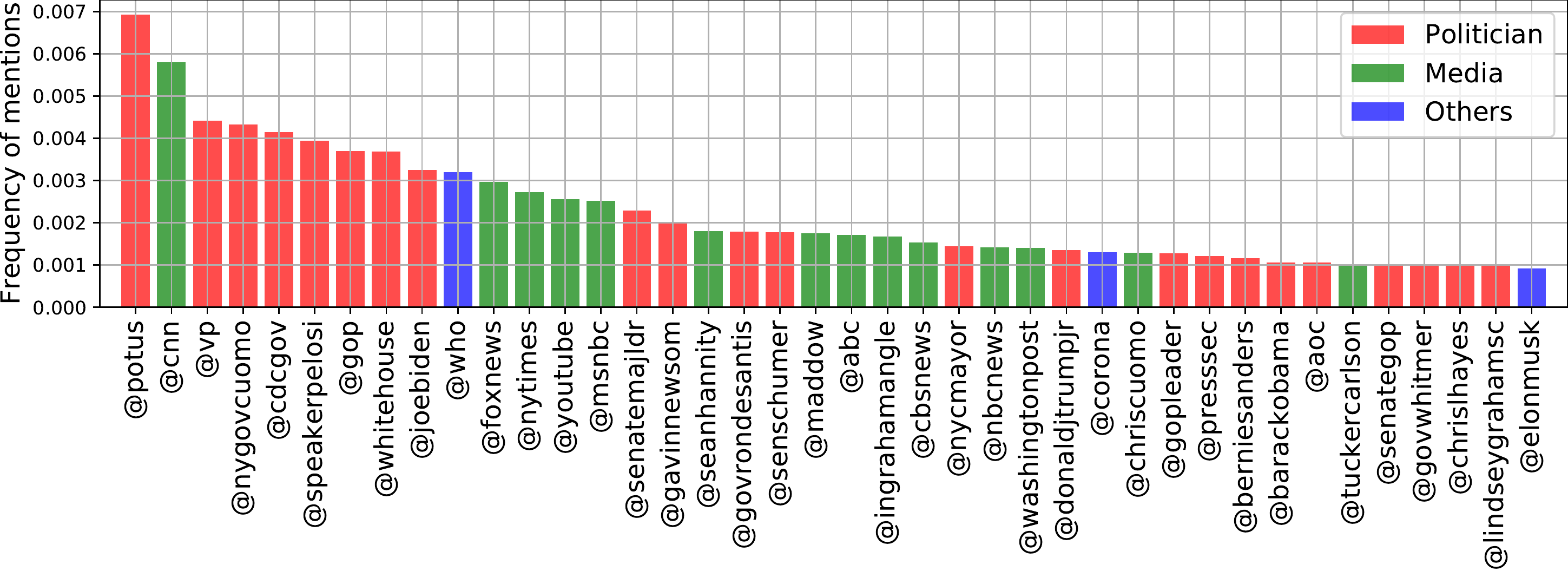}
  \caption{The 40 most frequently mentioned Twitter accounts. The most popular mention @realDonaldTrump (accounting for more than 4.5\%) are not displayed.
  \label{fig:mentioned}}
\end{figure}

People use @mentions to get someone's attention on social networks.
We found most of the frequent mentions were about politicians and news media, as illustrated in Figure~\ref{fig:mentioned}.
The mention of @realDonaldTrump accounted for 4.5\% of all mentions and was the most popular one.
To make Figure~\ref{fig:mentioned} more readable, the mention of @realDonaldTrump was not plotted.
Other national (e.g., @VP, and @JeoBiden) and regional (e.g., @NYGovCuomo, and @GavinNewsom) politicians were mentioned many times.
As news channels played a crucial role in broadcasting the updated COVID-19 news and policies to the public, it is not surprising to observe news media such as @CNN, @FoxNews, @nytimes, and @YouTube are prevalent in Figure~\ref{fig:mentioned}.
In addition, the World Health Organization @WHO, the beer brand @corona, and Elon Musk @elonmusk were among the top 40 mentions. 

\subsection{Topic Modeling}

To further explore what people tweeted, we adopted latent Dirichlet allocation (LDA)~\cite{blei2003latent} to infer coherent topics from plain-text tweets.
We created a tweet corpus by treating each unique tweet as one document.
Commonly used text preprocessing techniques, such as tokenization, lemmatization, and removing stop words, were then applied on each document to improve modeling performance.
Next, we performed the term frequency-inverse document frequency (TF-IDF) on the whole tweet corpus to assign higher weights to most import words.
Finally, the LDA model was applied on the TF-IDF corpus to extract latent topics.

We determined the optimal number of topics in LDA using $C_v$ metric, which was reported as the best coherence measure by combining normalized pointwise mutual information (NPMI) and the cosine similarity~\cite{roder2015exploring}.
For each topic number, we trained 500-pass LDA models for ten times.
We found the average $C_v$ scores demonstrated an increasing trend as the topic number became larger.
But the increasing speed became relatively slow if more than ten topics were considered.
Therefore, we chose ten as the most suitable topic number in our study.

The ten topics and words in each topic are illustrated in Table~\ref{tab:topic_model}.
We can see that Topic 1 is mostly related to statistics of COVID-19, such as deaths, cases, tests, and rates.
In the topic of treatment, healthcare related words, e.g., ``mask'', ``patient'', ``hospital'', ``nurse'', ``medical'', and ``PPE'', are clustered together.
Topic 3 is about politics, as top keywords include ``Trump'', ``president'', ``vote'', and ``democratic''.
The emotion topic mainly consists of informal language expressing emotions. 
Topic 5 is related to impact on work, businesses, and schools.
We believe Americans who are bilingual in Spanish and English contributed to the topic of Spanish.
Topic 7 is calling for unity in the community.
In the topic of places, many states (e.g., Florida and California) and cities (e.g., New York and San Francisco) are mentioned.
The last two topics are about praying and home activities when people followed stay-at-home orders. 
These topics are very informative and well summarize the overall conversations on social media.

\begin{table}[h]
\setlength{\tabcolsep}{2pt}
\centering
\caption{Top 40 keywords for the 10 topics extracted using the LDA topic model}\label{tab:topic_model}
\begin{tabular}{@{}ccccccccccc@{}}
\toprule
     & Topic 1             & Topic 2            & Topic 3            & Topic 4          & Topic 5                & Topic 6          & Topic 7            & Topic 8        & Topic 9                  & Topic 10                   \\ 
Rank & \textbf{Facts} & \textbf{Healthcare} & \textbf{Politics} & \textbf{Emotion} & \textbf{Business} & \textbf{Spanish} &  \textbf{Community} & \textbf{Location} & \textbf{Praying} & \textbf{Activities} \\ 
\midrule
1    & death               & mask               & trump              & viru             & money                  & que              & help               & case           & famili                   & quarantin                  \\
2    & test                & wear               & presid             & shit             & pay                    & lo               & amp                & counti         & stay                     & day                        \\
3    & flu                 & patient            & peopl              & peopl            & bill                   & por              & thank              & new            & love                     & time                       \\
4    & viru                & hospit             & viru               & fuck             & busi                   & del              & commun             & test           & god                      & amp                        \\
5    & peopl               & nurs               & american           & got              & fund                   & la               & work               & state          & thank                    & today                      \\
6    & case                & worker             & lie                & go               & need                   & para             & need               & confirm        & home                     & new                        \\
7    & number              & medic              & amp                & realli           & unemploy               & con              & health             & york           & friend                   & game                       \\
8    & china               & doctor             & nt                 & think            & work                   & una              & student            & posit          & one                      & play                       \\
9    & die                 & disinfect          & vote               & know             & tax                    & angel            & support            & order          & pray                     & watch                      \\
10   & rate                & healthcar          & say                & hand             & help                   & como             & great              & san            & pleas                    & music                      \\
11   & infect              & inject             & democrat           & gon              & stimulu                & pero             & time               & citi           & safe                     & home                       \\
12   & vaccin              & ppe                & china              & thing            & relief                 & est              & pleas              & updat          & amp                      & season                     \\
13   & spread              & face               & respons            & get              & compani                & vega             & pandem             & close          & time                     & one                        \\
14   & say                 & protect            & america            & one              & peopl                  & hay              & inform             & death          & bless                    & sport                      \\
15   & report              & front              & countri            & nt               & amp                    & su               & school             & via            & life                     & love                       \\
16   & know                & line               & know               & want             & job                    & son              & share              & home           & prayer                   & go                         \\
17   & mani                & treat              & call               & home             & worker                 & no               & learn              & florida        & day                      & movi                       \\
18   & diseas              & drug               & news               & time             & due                    & persona          & join               & governor       & hope                     & make                       \\
19   & symptom             & care               & need               & stay             & via                    & esto             & import             & beach          & good                     & see                        \\
20   & million             & thank              & think              & make             & million                & solo             & crisi              & report         & happi                    & look                       \\
21   & one                 & use                & want               & take             & market                 & sin              & donat              & california     & peopl                    & thank                      \\
22   & amp                 & hero               & right              & back             & pandem                 & casa             & impact             & reopen         & know                     & due                        \\
23   & countri             & work               & pandem             & work             & small                  & ser              & provid             & break          & work                     & social                     \\
24   & new                 & ventil             & one                & day              & stock                  & le               & read               & health         & lost                     & distanc                    \\
25   & day                 & help               & blame              & say              & insur                  & mundo            & resourc            & open           & today                    & got                        \\
26   & year                & need               & medium             & as               & govern                 & puerto           & team               & francisco      & help                     & viru                       \\
27   & popul               & fight              & state              & lol              & food                   & hoy              & take               & stay           & die                      & good                       \\
28   & nt                  & glove              & polit              & need             & make                   & rico             & public             & announc        & year                     & night                      \\
29   & think               & treatment          & die                & wash             & dollar                 & do               & make               & today          & everyon                  & pandem                     \\
30   & kill                & frontlin           & republican         & still            & employe                & eso              & social             & texa           & take                     & show                       \\
31   & state               & amp                & hoax               & come             & paid                   & sea              & respons            & first          & keep                     & year                       \\
32   & world               & via                & tri                & see              & check                  & dio              & continu            & day            & see                      & video                      \\
33   & data                & respond            & make               & even             & price                  & han              & busi               & resid          & heart                    & walk                       \\
34   & cdc                 & trump              & cnn                & feel             & trump                  & california       & updat              & gov            & mom                      & back                       \\
35   & immun               & bleach             & time               & stop             & state                  & gracia           & onlin              & total          & viru                     & week                       \\
36   & caus                & lysol              & stop               & sick             & give                   & ma               & care               & offici         & best                     & store                      \\
37   & time                & suppli             & elect              & damn             & rent                   & nada             & proud              & say            & posit                    & cancel                     \\
38   & even                & test               & go                 & everyon          & feder                  & pod              & today              & school         & need                     & fan                        \\
39   & lab                 & clinic             & white              & someon           & american               & muy              & educ               & mayor          & feel                     & fun                        \\
40   & studi               & cure               & take               & said             & cut                    & tan              & local              & area           & go                       & work                       \\ \bottomrule
\end{tabular}
\end{table}

\section{Sentiment Analysis}

In this section, we conducted a comprehensive sentiment analysis from three aspects.
First, the overall public emotions were investigated using polarized words and facial emojis.
Then, we studied how sentiment changed over time at the national and state levels during the COVID-19 pandemic. 
Finally, event-specific emotions were reported.



\subsection{Emotionally Polarized Tweets}
Emotionally polarized words or sentences express either positive or negative emotions.
We leveraged TextBlob~\cite{loria2014textblob} to estimate the sentimental polarity of words and tweets.
For each word, TextBlob offers a subjectivity score within the range [0.0, 1.0] where 0.0 is most objective and 1.0 is most subjective, and a polarity score within the range [-1.0, 1.0] where -0.1 is the most negative and 1.0 is the most positive.
We used the subjectivity threshold to filter out objective tweets, and used the polarity threshold to determine the sentiment.
For example, a subjectivity threshold of 0.5 would only select the tweets with a subjectivity score greater than 0.5 as the candidates for polarity checking.
A polarity threshold of 0.7 treated tweets with a polarity score greater than 0.7 as positive and those with a polarity less than -0.7 as negative.

\begin{figure}[h]
\centering
  \includegraphics[width=0.7\linewidth]{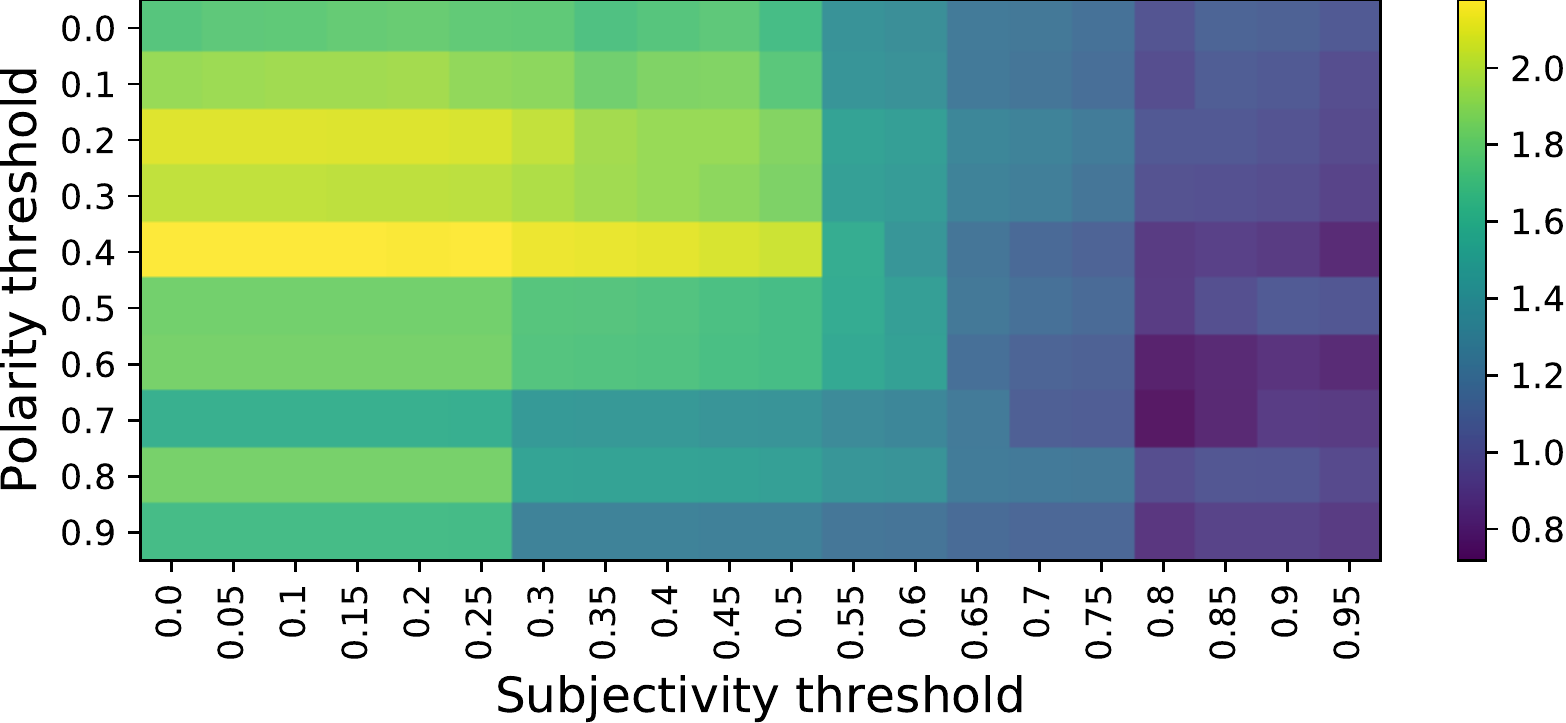}
  \caption{The ratio of \# of positive tweets over \# of negative tweets with different polarity and subjectivity thresholds. Positive emotions dominate when the ratio is greater than one. Otherwise, negative emotions are more popular.
  \label{fig:textblob_heatmap}}
\end{figure}

\begin{figure}[h]
  \centering
  \subfigure[Ratio > 1
  \label{fig:wordcloud_greater_1}]{\includegraphics[width=0.32\linewidth]{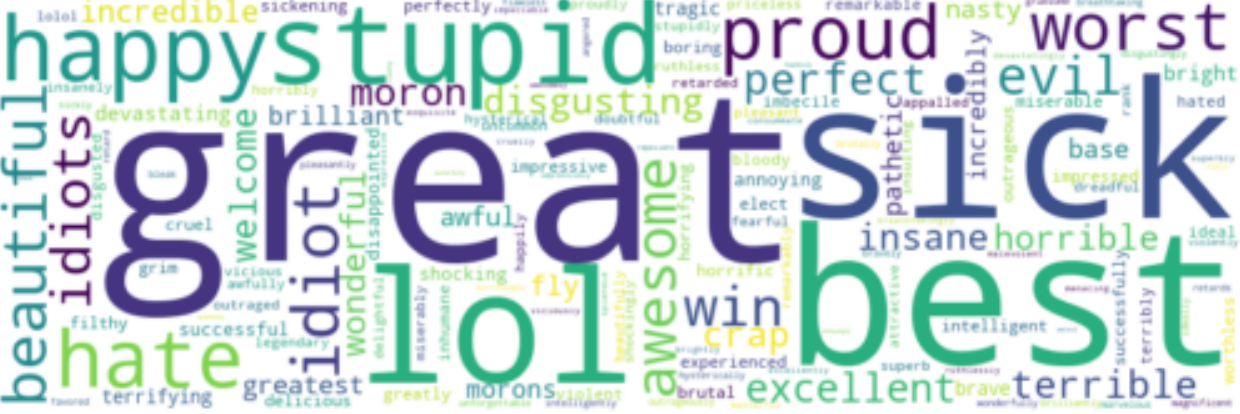}}
  \subfigure[Ratio $\approx$ 1
  \label{fig:wordcloud_equal_1}]{\includegraphics[width=0.32\linewidth]{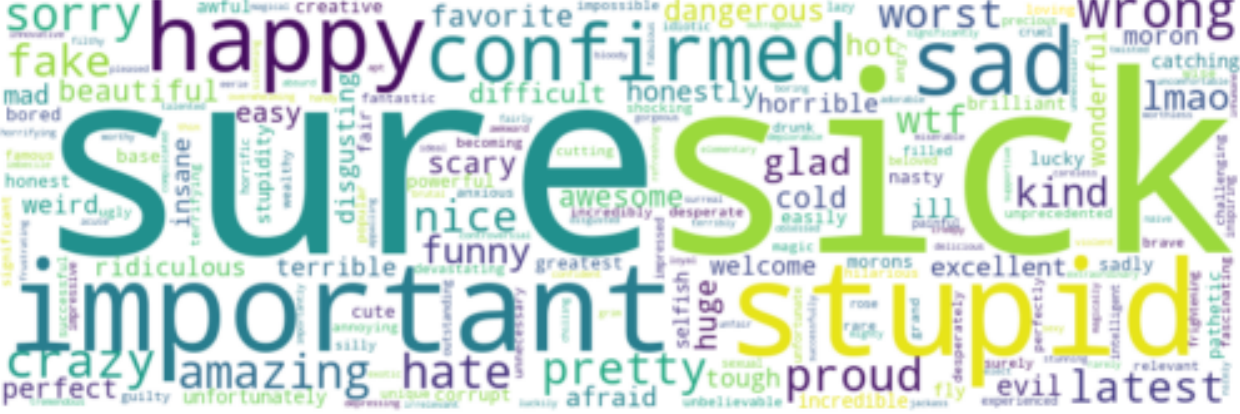}}
  \subfigure[Ratio < 1
  \label{fig:wordcloud_less_1}]{\includegraphics[width=0.32\linewidth]{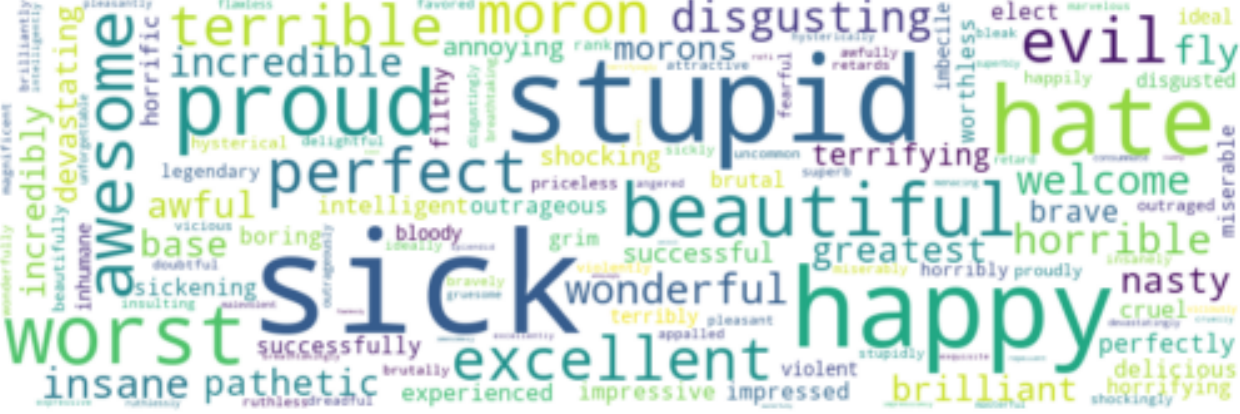}}
  \caption{Polarized word clouds of different positive/negative ratios. (a) was generated with thresholds (subjectivity=0.2, polarity=0.7), (b) with (subjectivity=0.8, polarity=0.2), and (c) with (subjectivity=0.8, polarity=0.7).
  \label{fig:word_cloud}}
\end{figure}

Figure~\ref{fig:textblob_heatmap} illustrates the ratio of the number of positive tweets over negative ones with different combinations of subjectivity and polarity thresholds.
Positive and negative emotions evenly matched with each other when the ratio equaled one.
We can see that emotion patterns changes along with threshold settings. 
Specifically, positive emotions dominated on Twitter with small polarity and subjectivity thresholds.
However, negative emotions became to overshadow the positive ones under large polarity and subjectivity thresholds.
Figure~\ref{fig:word_cloud} shows three examples of polarized word clouds where the ratio was greater than 1 (subjectivity=0.2, polarity=0.7), equal to 1 (subjectivity=0.8, polarity=0.2), and less than 1 (subjectivity=0.8, polarity=0.7).

\subsection{Facial Emoji Patterns}

Besides polarized-text based sentiment analysis, we took advantage of facial emojis to further study the public emotions.
Facial emojis are suitable to measure tweet sentiments because they are ubiquitous on social media, conveying diverse positive, neutral, and negative feelings.
We grouped the sub-categories of facial emojis suggested by the Unicode Consortium into positive, neutral, and negative categories.
Specifically, all face-smiling, face-affection, face-tongue, face-hat emojis, and~\includegraphics[width=0.02\linewidth]{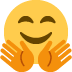}~\includegraphics[width=0.02\linewidth]{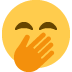}~\includegraphics[width=0.02\linewidth]{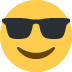}~\includegraphics[width=0.02\linewidth]{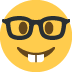} were regarded as positive; all face-neutral-skeptical, face-glasses emojis, and ~\includegraphics[width=0.02\linewidth]{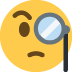}~\includegraphics[width=0.02\linewidth]{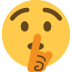}~\includegraphics[width=0.02\linewidth]{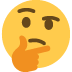} were grouped as neutral; and all face-sleepy, face-unwell, face-concerned, face-negative emojis were treated as negative. A full list of our emoji emotion categories are available at \url{http://covid19research.site/emoji-category/}.

We detected 4,739 (35.2\%) positive emojis, 2,438 (18.1\%) neutral emojis, and 6,271 (46.6\%) negative emojis in our dataset.
Negative emojis accounted for almost half of all emoji usages. 
Table~\ref{tab:emojis} illustrates top emojis by sentiment categories with their usage frequencies.
The most frequent emojis in the three categories were very representative.
As expected,~\includegraphics[width=0.02\linewidth]{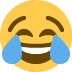} still was the most popular emojis in all categories, which kept consistent with many other recent research findings\cite{meaning-of-face-with-tears-of-joy,feng2019chasing}.
The thinking face emoji~\includegraphics[width=0.02\linewidth]{figures/emoji_image/1f914.png} was the most widely used neutral facial emoji, indicting people were puzzled on COVID-19.
Surprisingly, the face with medical mask emoji~\includegraphics[width=0.02\linewidth]{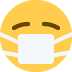} ranked higher than any other negative emojis.
The skull emoji~\includegraphics[width=0.02\linewidth]{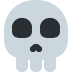} appeared more frequently than any other positive and neutral emojis except~\includegraphics[width=0.02\linewidth]{figures/emoji_image/1f602.png}~\includegraphics[width=0.02\linewidth]{figures/emoji_image/1f914.png}~\includegraphics[width=0.02\linewidth]{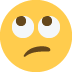}~\includegraphics[width=0.02\linewidth]{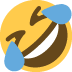}.
We think the sneezing face~\includegraphics[width=0.02\linewidth]{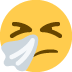} and the hot face emoji~\includegraphics[width=0.02\linewidth]{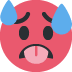} are very likely to be relative to suspected symptoms of COVID-19.

\begin{table}[h] 
\setlength{\tabcolsep}{4.4pt}
\centering 
\caption{Top emojis by sentiment categories (numbers represent frequency)}\label{tab:emojis} 
 \begin{tabular}{lcccccccccccccccccccccccccccccc} 
 \toprule
Positive & \includegraphics[width=0.05\linewidth]{figures/emoji_image/1f602.png} & \includegraphics[width=0.05\linewidth]{figures/emoji_image/1f923.png} & \includegraphics[width=0.05\linewidth]{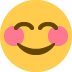} & \includegraphics[width=0.05\linewidth]{figures/emoji_image/1f60e.png} & \includegraphics[width=0.05\linewidth]{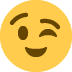} & \includegraphics[width=0.05\linewidth]{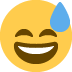} & \includegraphics[width=0.05\linewidth]{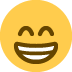} & \includegraphics[width=0.05\linewidth]{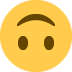} & \includegraphics[width=0.05\linewidth]{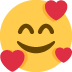} & \includegraphics[width=0.05\linewidth]{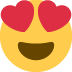} & \includegraphics[width=0.05\linewidth]{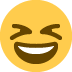} & \includegraphics[width=0.05\linewidth]{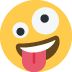} &  \includegraphics[width=0.05\linewidth]{figures/emoji_image/1f917.png} & \\
& 1608 & 522 & 199 & 189 & 176 & 173 & 171 & 162 & 153 & 150 & 139 & 138 & 132 \\
& \includegraphics[width=0.05\linewidth]{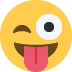} & \includegraphics[width=0.05\linewidth]{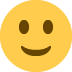} & \includegraphics[width=0.05\linewidth]{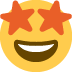} & \includegraphics[width=0.05\linewidth]{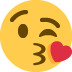} & \includegraphics[width=0.05\linewidth]{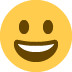} & \includegraphics[width=0.05\linewidth]{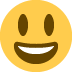} & \includegraphics[width=0.05\linewidth]{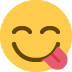} & \includegraphics[width=0.05\linewidth]{figures/emoji_image/1f913.png} & \includegraphics[width=0.05\linewidth]{figures/emoji_image/1f92d.png} & \includegraphics[width=0.05\linewidth]{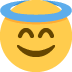} & \includegraphics[width=0.05\linewidth]{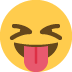} & \includegraphics[width=0.05\linewidth]{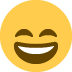} & \includegraphics[width=0.05\linewidth]{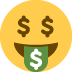} & \\ 
& 95 & 91 & 77 & 77 & 67 & 59 & 56 & 52 & 50 & 49 & 44 & 28 & 27 \\ \midrule 
Neutral & \includegraphics[width=0.05\linewidth]{figures/emoji_image/1f914.png} & \includegraphics[width=0.05\linewidth]{figures/emoji_image/1f644.png} & \includegraphics[width=0.05\linewidth]{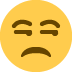} & \includegraphics[width=0.05\linewidth]{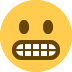} & \includegraphics[width=0.05\linewidth]{figures/emoji_image/1f9d0.png} & \includegraphics[width=0.05\linewidth]{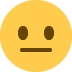} & \includegraphics[width=0.05\linewidth]{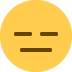} & \includegraphics[width=0.05\linewidth]{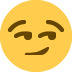} & \includegraphics[width=0.05\linewidth]{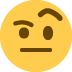} & \includegraphics[width=0.05\linewidth]{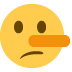} & \includegraphics[width=0.05\linewidth]{figures/emoji_image/1f92b.png} & \includegraphics[width=0.05\linewidth]{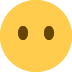} & \includegraphics[width=0.05\linewidth]{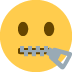}  \\ 
&  865 & 591 & 197 & 146 & 143 & 104 & 104 & 87 & 81 & 43 & 36 & 27 & 14 \\ \midrule 
Negative & \includegraphics[width=0.05\linewidth]{figures/emoji_image/1f637.png} & \includegraphics[width=0.05\linewidth]{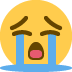} & \includegraphics[width=0.05\linewidth]{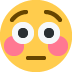} & \includegraphics[width=0.05\linewidth]{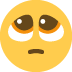} & \includegraphics[width=0.05\linewidth]{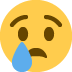} & \includegraphics[width=0.05\linewidth]{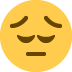} & \includegraphics[width=0.05\linewidth]{figures/emoji_image/1f480.png} & \includegraphics[width=0.05\linewidth]{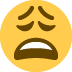} & \includegraphics[width=0.05\linewidth]{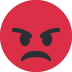} & \includegraphics[width=0.05\linewidth]{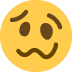} & \includegraphics[width=0.05\linewidth]{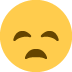} & \includegraphics[width=0.05\linewidth]{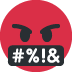} & \includegraphics[width=0.05\linewidth]{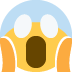} & \\
& 1167 & 629 & 446 & 401 & 394 & 368 & 251 & 245 & 227 & 207 & 184 & 159 & 145 \\
& \includegraphics[width=0.05\linewidth]{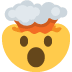} & \includegraphics[width=0.05\linewidth]{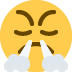} & \includegraphics[width=0.05\linewidth]{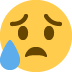} & \includegraphics[width=0.05\linewidth]{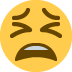} & \includegraphics[width=0.05\linewidth]{figures/emoji_image/1f927.png} & \includegraphics[width=0.05\linewidth]{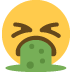} & \includegraphics[width=0.05\linewidth]{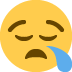} & \includegraphics[width=0.05\linewidth]{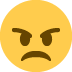} & \includegraphics[width=0.05\linewidth]{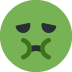} & \includegraphics[width=0.05\linewidth]{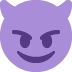} & \includegraphics[width=0.05\linewidth]{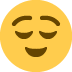} & \includegraphics[width=0.05\linewidth]{figures/emoji_image/1f975.png} & \includegraphics[width=0.05\linewidth]{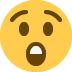} &  \\ 
& 117 & 104 & 102 & 95 & 75 & 70 & 70 & 68 & 68 & 63 & 59 & 54 & 47 \\
\bottomrule
\end{tabular} 
 \end{table}
 
\subsection{Sentiment Over Time}
\begin{figure}[h]
  \subfigure[Daily overall sentiment aggregated by all states
  \label{fig:distribution_by_day_by_emotion}]{\includegraphics[width=\linewidth]{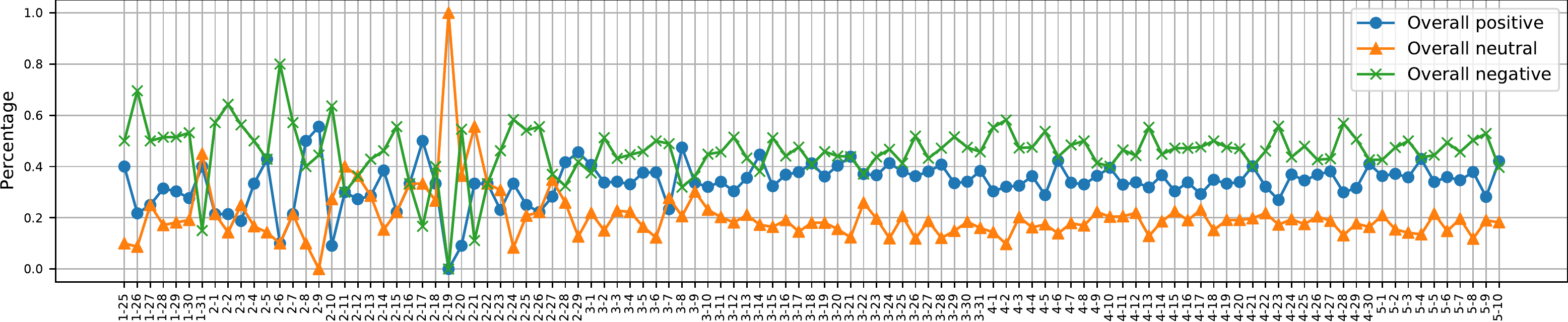}}
  
  \subfigure[Daily positive sentiment by state
  \label{fig:distribution_by_day_by_state_face_positive}]{\includegraphics[width=\linewidth]{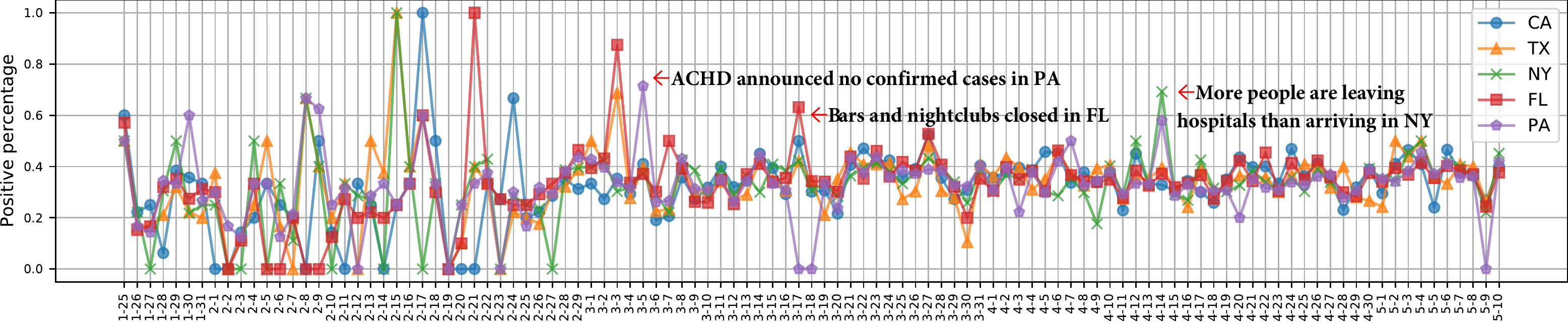}}
  
  \subfigure[Daily neutral sentiment by state
  \label{fig:distribution_by_day_by_state_face_neutral}]{\includegraphics[width=\linewidth]{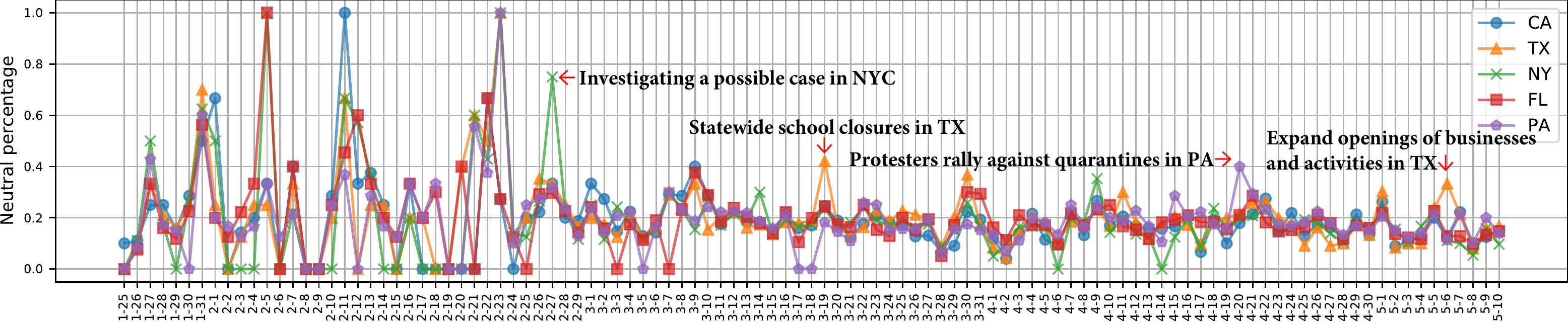}}
  
  \subfigure[Daily negative sentiment by state
  \label{fig:distribution_by_day_by_state_face_negative}]{\includegraphics[width=\linewidth]{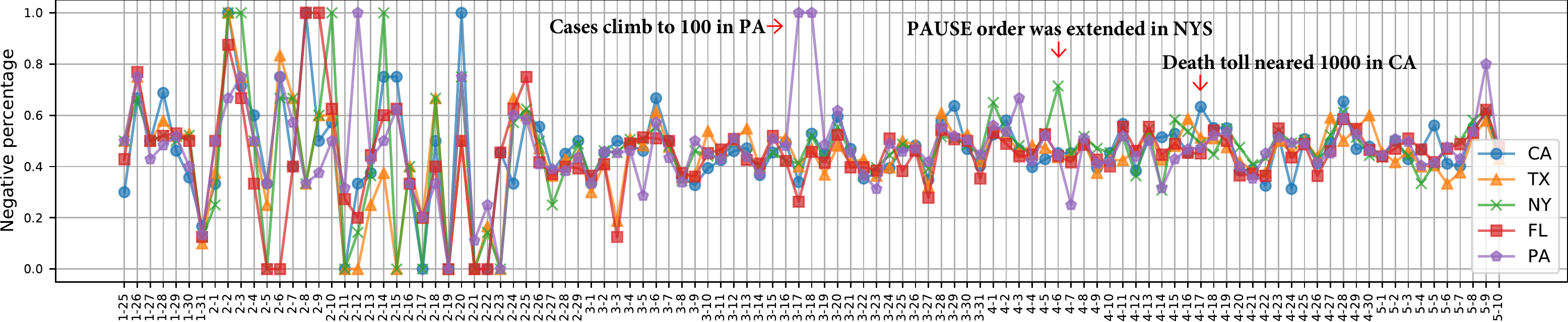}}
  \caption{Emotion distribution by day
  \label{fig:sentiment_by_day}}
\end{figure}

We used facial emojis to track the different types of public sentiment during the COVID-19 pandemic.
Figure~\ref{fig:distribution_by_day_by_emotion} shows the daily overall emotions aggregated by all states.
In Phase 1 (from Jan. 25 to Feb. 24), the publish emotions changed in large ranges due to the data sparsity.
In Phase 2 (from Feb. 25 to Mar. 14), positive and negative emotions overshadowed each other dynamically but demonstrated stable trends.
In Phase 3 (from Mar. 15 to May 10), negative sentiment dominated both positive and neutral emotions, expressing the public's concerns on COVID-19.

We also investigated the daily positive, neutral, and negative sentiment of different states as presented in Figure~\ref{fig:distribution_by_day_by_state_face_positive}, Figure~\ref{fig:distribution_by_day_by_state_face_neutral}, and Figure~\ref{fig:distribution_by_day_by_state_face_negative} respectively.
The top five states with the highest tweet volumes, i.e., CA, TX, NY, FL, and PA, were taken as examples.
Similar to Phase 1 patterns in Figure~\ref{fig:distribution_by_day_by_emotion}, the expression of emotion by people in different states varied greatly.
In Phase 2 and Phase 3, the five states demonstrated similar positive, neutral, and negative patterns at most dates, as their sentiment percentages were cluttered together and even overlapped.
However, there existed state-specific emotion outliers in Phase 2 and Phase 3.
For example, the positive sentiment went up to 70\% in PA when the Allegheny County Health Department (ACHD) announced there were no confirmed cases of COVID-19 in Pennsylvania on Mar. 5.
People in New York state expressed more than 75\% neutral sentiments on Feb. 27 when the New York City Health Department announced that it was investigating a possible COVID-19 case in the city.
On Mar. 17 and Mar. 18, residents in PA demonstrated almost 100\% negative sentiments when the statewide COVID-19 confirmed cases climbed to 100.

\subsection{Event-specific Sentiment\label{sec:eventsent}} 
\begin{figure}[h]
\centering
  \includegraphics[width=\linewidth]{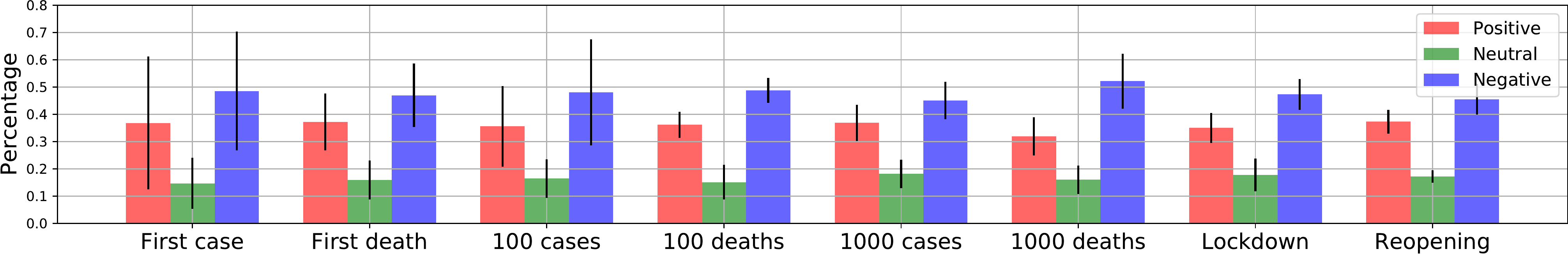}
  \caption{Even-specific sentiments. The means of positive, neutral, and negative emotions are very close but with different standard deviations.
  \label{fig:event_specific_sentiment}}
\end{figure}

We studied the event-specific sentiment by aggregating tweets posted from different states when the same critical COVID-19 events occurred.
We focused on the following eight events:
\begin{itemize}
    \item The first, the $100^{th}$, and the $1000^{th}$ confirmed COVID-19 cases, 
    \item The first, the $100^{th}$, and the $1000^{th}$ confirmed COVID-19 deaths, 
    \item Lockdown 
    \item Reopen
\end{itemize}
For the first seven events, we aggregated the tweets in CA, TX, FL, NY, GA, PA, IL, MD, VA, and AZ, which were also studied in Subsection~\ref{sec:stay-at-home-work-engagement}.
For the last event, we investigated the nine states of TX, GA, TN, CO, AL, MS, ID, AK, and MT, which kept consistent with Subsection~\ref{sec:stay-at-home-work-engagement}.
To our surprise, the average percentages of each sentiment type in the eight events demonstrated similar patterns, as shown in Figure~\ref{fig:event_specific_sentiment}.
We carried out the one-way multivariate analysis of variance (MANOVA) and found the p-value was nearly 1.0, indicating there was no significant difference among these eight event-specific sentiments.
When the first case, $100^{th}$ cases, and first death were confirmed, sentiment standard deviations were much larger than the rest events, suggesting people in different states expressed varying and diverse sentiments at the beginning of COVID-19 outbreak.
The negative emotion reached the highest level among all events when $1000^{th}$ deaths were reported.
The positive emotion achieved the highest level among all events when states began to reopen.

\section{Discussions and Conclusion}

This paper presents a large public geo-tagged COVID-19 Twitter dataset containing 650,563 unique geo-tagged COVID-19 tweets posted in the United States from Jan. 25 to May 10.
A small number of tweets were missing during the data collection period due to corrupted files and intermittent internet connectivity issues.
We compensated for the data gaps using the COVID-19 dataset collected by Chen et al.~\cite{chen2020covid}.
As different COVID-19 keywords were used in~\cite{chen2020covid} and our study to filter tweet streaming, it did not compensate for the missing data perfectly.
However, given the small proportion of missing data, we do not expect the conclusions to change.
For more details about our dataset, please refer to Appendix~\ref{dataset}.

Based on the geo-tagged dataset, we investigated fine-grained public reactions during the COVID-19 pandemic.
First, we studied the daily tweeting patterns in different states and found most state pairs had a strong linear correlation.
The local time zones inferred from tweet locations make it possible to compare the hourly tweeting behaviors on workdays and weekends.
Their different hourly patterns during 8:00 to 17:00 inspired us to propose approaches to measure work engagement.
Second, we utilized tweet locations to explore geographic distributions of COVID-19 tweets at state and county levels.
Third, we summarized and revealed the themes people discussed on Twitter using both social network exclusive tools (i.e., \#hashtags and @mentions) and general text-based topic models.
Finally, we reported comprehensive sentiment analytics, including the overall public emotions, how public feelings changed over time, and the expressed emotions when specific events occurred.
Hopefully, this geo-tagged Twitter dataset can facilitate more fine-grained COVID-19 studies in the future.

\begin{appendices}
    \section{Dataset}
\label{dataset}


In this section, we first described how we collected Twitter data and compensated for data gaps.
Then we removed Twitter bots to enhance data analytics. 
At last, we extracted the U.S. geo-tagged COVID-19 tweets from general tweets.

\subsection{Data Collection}\label{sec:data_collection}

We utilized Twitter's Streaming APIs to crawl real-time tweets containing a set of ``coronavirus'', ``wuhan'', ``corona'', ``nCoV'' keywords related to the novel coronavirus outbreak since January 25, 2020.\footnote{This is two days after Wuhan lockdown.}
After the World Health Organization (WHO) announced the official name of COVID-19 on February 11, 2020, we added ``COVID19'', ``COVID\begin{CJK}{UTF8}{min}ー\end{CJK}19'', ``coronapocalypse'',  ``Coronavid19'', ``Covid\_19'',  ``COVID-19'', and ``covid'' into our keyword set.
We collected more than 170 million tweets generated by 2.7 million unique users from January 25 to May 10, 2020.
Each tweet was formatted in a JSON file with named attributes and associated values. 
We lost 38.5\% tweets uniformly distributed among May 18 and Apr. 4 due to corrupted files, and missed 88 hours of data because of intermittent internet connectivity issues in the entire data collection period.
More details about data gaps are available at \url{http://covid19research.site/geo-tagged_twitter_datasets/known_data_gaps.csv}.
To compensate for these data gaps, we sought for the COVID-19 dataset maintained by Chen et al.~\cite{chen2020covid} and downloaded 16,459,659 tweets.

\begin{figure}[h]
  \includegraphics[width=\linewidth]{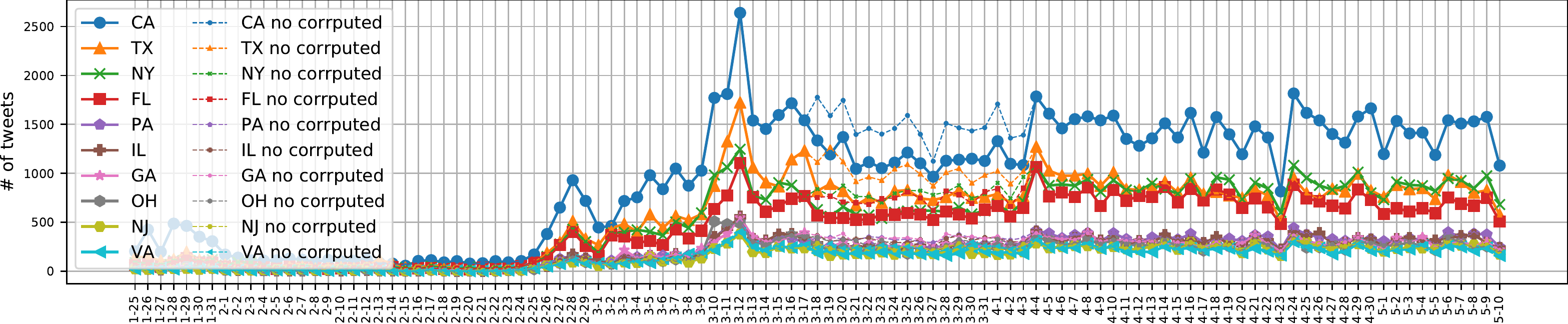}
  \caption{The daily number of tweets from the top 10 states generating most tweets. 
  \label{fig:daily_pattern}}
\end{figure}

\subsection{Data Cleaning}

One of the challenges when dealing with messy text like tweets is to remove noisy data generated by Twitter bots.
Inspired by the bot detection approach proposed in~\cite{ljubevsic2016global}, we conceived the two types of Twitter users as bots: (1) those who posted more than 5000 COVID-19 tweets (more than 46 tweets on average per day) during our data collection period; (2) those who posted over 1000 COVID-19 tweets in total and the top three frequent posting intervals covered at least their 90\% tweets.
For the two types of bots, we removed 317,101 tweets created by 32 bots and 120,932 tweets by 36 bots respectively.



\subsection{Geo-tagged Data in the U.S.}
\begin{figure}[ht]
  \includegraphics[width=\linewidth]{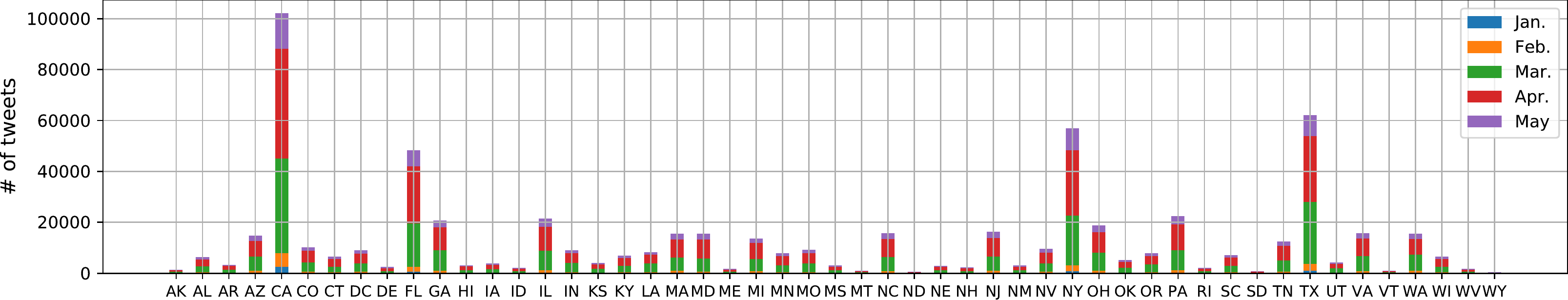}
  \caption{The monthly number of geo-tagged tweets in 50 states and Washington D.C. in the United States.
  \label{fig:monthly_tweet}}
\end{figure}

Twitter allows users to optionally tag tweets with different precise geographic information, indicating the real-time location of users when tweeting.
Typical tweet locations can be either a box polygon of coordinates specifying general areas like cities and neighborhoods, or an exact GPS latitude and longitude coordinate.
We detected and examined the ``place'' attribute in collected tweet JSON files.
If the embedded ``country\_code'' was ``US'' and the extracted state was among the 50 states and Washington D.C. in the United States, we added the tweet into our geo-tagged dataset.
After removing retweets, 650,563 geo-tagged unique tweets from 246,032 users in the United States were collected.
Among them, 38,818 tweets (5.96\% of our dataset) were retrieved from the dataset proposed by Chen et al.~\cite{chen2020covid}.
The monthly number of geo-tagged tweets in each state is shown in Figure~\ref{fig:monthly_tweet}.

\subsection{Twitter User Analysis}
\begin{figure}[ht]
  \centering
  \includegraphics[width=0.5\linewidth]{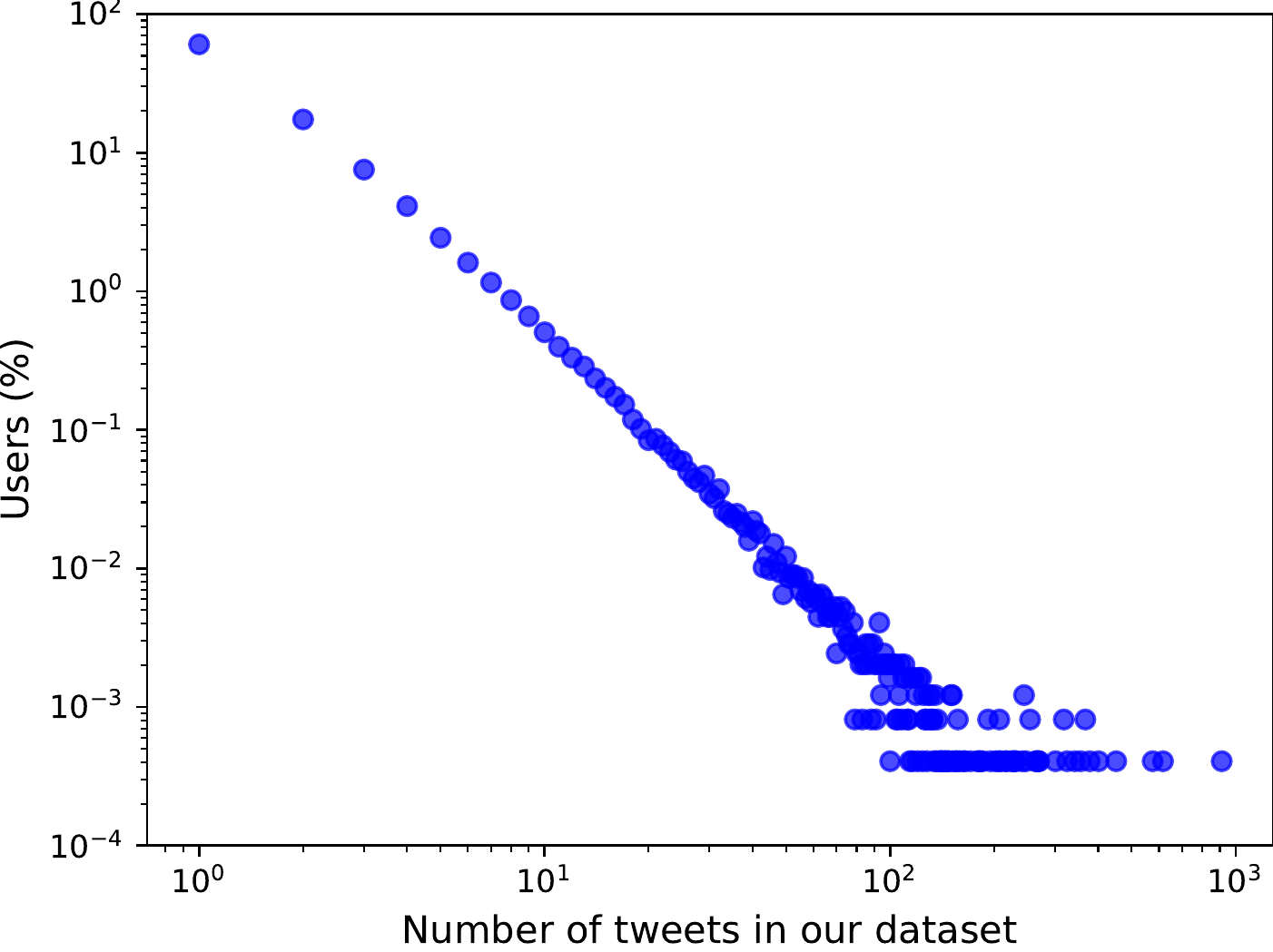}
  \caption{Scatter plot of number of tweets per user on log-log scale.
  \label{fig:user_analysis}}
\end{figure}

We further analyzed the users in our dataset to demonstrate they crowdsourced the public. 
Figure~\ref{fig:user_analysis} shows the user proportion versus the number of posted tweets.
We found only 0.055\% users tweeted more than one geo-tagged tweet per day on average, generating 11,844 tweets (1.82\% of all tweets) in our dataset.
To be specific, 96.71\% users had no more than ten records in our dataset.

\end{appendices}
    

\bibliographystyle{unsrt} 
\bibliography{references}  

\begin{thebibliography}{10}

\bibitem{feb25warning}
E.J. MUNDELL and Robin Foster.
\newblock {\em Coronavirus Outbreak in America Is Coming: CDC}, 2020.

\bibitem{mar13emergency}
The~White House.
\newblock {\em Proclamation on Declaring a National Emergency Concerning the
  Novel Coronavirus Disease (COVID-19) Outbreak}, 2020.

\bibitem{Nominatim}
OpenStreetMap.
\newblock {\em Nominatim: a search engine for OpenStreetMap data}, 2018.

\bibitem{lockdownWiki}
Wikipedia.
\newblock {\em U.S. state and local government response to the COVID-19
  pandemic}, 2020.

\bibitem{ReopenNYT}
Sarah Mervosh, Jasmine~C. Lee, Lazaro Gamio, and Nadja Popovich.
\newblock {\em Coronavirus Outbreak in America Is Coming: CDC}, 2020.

\bibitem{blei2003latent}
David~M Blei, Andrew~Y Ng, and Michael~I Jordan.
\newblock Latent dirichlet allocation.
\newblock {\em Journal of machine Learning research}, 3(Jan):993--1022, 2003.

\bibitem{roder2015exploring}
Michael R{\"o}der, Andreas Both, and Alexander Hinneburg.
\newblock Exploring the space of topic coherence measures.
\newblock In {\em Proceedings of the eighth ACM international conference on Web
  search and data mining}, pages 399--408. ACM, 2015.

\bibitem{loria2014textblob}
Steven Loria, P~Keen, M~Honnibal, R~Yankovsky, D~Karesh, E~Dempsey, et~al.
\newblock Textblob: simplified text processing.
\newblock {\em Secondary TextBlob: Simplified Text Processing}, 2014.

\bibitem{meaning-of-face-with-tears-of-joy}
Dictionary.com.
\newblock {\em What does the face with tears of joy emoji mean}, 2017.

\bibitem{feng2019chasing}
Yunhe Feng, Zheng Lu, Zhonghua Zheng, Peng Sun, Wenjun Zhou, Ran Huang, and
  Qing Cao.
\newblock Chasing total solar eclipses on twitter: Big social data analytics
  for once-in-a-lifetime events.
\newblock In {\em 2019 IEEE Global Communications Conference (GLOBECOM)}, pages
  1--6. IEEE, 2019.

\bibitem{chen2020covid}
Emily Chen, Kristina Lerman, and Emilio Ferrara.
\newblock Covid-19: The first public coronavirus twitter dataset.
\newblock {\em arXiv preprint arXiv:2003.07372}, 2020.

\bibitem{ljubevsic2016global}
Nikola Ljube{\v{s}}i{\'c} and Darja Fi{\v{s}}er.
\newblock A global analysis of emoji usage.
\newblock In {\em Proceedings of the 10th Web as Corpus Workshop}, pages
  82--89, 2016.

\end{thebibliography}

\end{document}